\documentclass[12pt,preprint,dvips]{aastex}
\usepackage{natbib}

\def \imsfrs{IM~SFRs}
\def \imsfr{IM~SFR}

\def \iras{\emph{IRAS}}
\def \msun{$\mathrm{M}_\odot$}

\def \lsun{$\mathrm{L}_\odot$}

\def \kms{km~s$^{-1}$}
\def \deg{$^\circ$}
\def \av{$A_{\mathrm{V}}$}
\def \ak{$A_{\mathrm{K}}$}
\def \ang{\AA}
\def \ks{$K_{S}$}
\def \jhks{\emph{$JHK_{S}$}}
\makeatletter
\newcommand{\Rom}[1]{\expandafter\@slowromancap\romannumeral #1@}
\makeatother

\author{Michael~J.~Lundquist\altaffilmark{1}}
\author{Henry~A.~Kobulnicky\altaffilmark{1}}
\author{Michael~J.~Alexander\altaffilmark{1,2}}
\author{Charles~R.~Kerton\altaffilmark{3}}
\author{Kim~Arvidsson\altaffilmark{4}}

\altaffiltext{1}{Department of Physics and Astronomy, University of Wyoming, Laramie, WY 82071 USA}
\altaffiltext{2}{Department of Physics, Lehigh University, 16 Memorial Drive East, Bethlehem, PA 18015 USA}
\altaffiltext{3}{Department of Physics and Astronomy, Iowa State University, Ames, IA 50011 USA}
\altaffiltext{4}{Trull School of Sciences \& Mathematics, Schreiner University, 2100 Memorial Blvd., Kerrville, TX 78028 USA}

\begin{document}

\title{AN ALL-SKY SAMPLE OF INTERMEDIATE-MASS STAR-FORMING REGIONS }

\begin{abstract}
We present an all-sky sample of 984 candidate intermediate-mass Galactic star-forming regions color-selected from the \emph{Infrared Astronomical Satellite} (\emph{IRAS}) Point Source Catalog and morphologically classify each object using mid-infrared \emph{Wide-field Infrared Survey Explorer} (\emph{WISE}) images.  Of the 984 candidates, 616 are probable star-forming regions (62.6\%), 128 are filamentary structures (13.0\%), 39 are point-like objects of unknown nature (4.0\%), and 201 are galaxies (20.4\%).  We conduct a study of four of these regions, IRAS~00259+5625, IRAS~00420+5530, IRAS~01080+5717, and IRAS~05380+2020, at Galactic latitudes $|b|$ $>$ 5\deg\ using  optical spectroscopy from the Wyoming Infrared Observatory along with near-infrared photometry from the Two-Micron All Sky Survey to investigate their stellar content.  New optical spectra, color-magnitude diagrams, and color-color diagrams reveal their extinctions, spectrophotometric distances, and the presence of small stellar clusters containing 20--78 \msun\ of stars.  These low-mass diffuse star clusters contain $\sim$65--250 stars for a typical initial mass function, including one or more mid-B stars as their most massive constituents. Using infrared spectral energy distributions we identify  young stellar objects near each region and assign probable masses and evolutionary stages to the protostars.  The total infrared luminosity lies in the range 190 to 960 \lsun, consistent with the sum of the luminosities of the individually identified young stellar objects.

\end{abstract}

\section{Introduction \label{intro.sec}}
It is well established that most stars form in groups or clusters \citep{Larson1982}.  \citet{Weidner2006} suggested that an underlying physical relationship exists between the most massive star in an embedded cluster and the total stellar mass of the cluster.  Low-mass stars tend to form in small, loose aggregates of 10--100 stars $\mathrm{pc}^{-3}$ \citep{Testi1999}, while high-mass stars are associated with clusters that contain up to 1000 stars $\mathrm{pc}^{-3}$ \citep{Hillenbrand1995}.  Evidence suggests that there may be a turn-on mass for stellar clustering at the scale of intermediate-mass stars of 8--10 \msun\ and that this onset may be sudden \citep{Testi1997} or gradual \citep{Testi1999}.  Hence, identifying and studying clusters having intermediate-mass stars as their most massive members may yield clues about the onset of cluster formation.  

While it is reasonably well understood how isolated, low-mass stars form \citep{Mckee2007}, it may not be as simple as scaling this standard theory to explain high-mass star formation \citep{Zinnecker2007}.  Scaling up low-mass star formation to higher masses ($>$10 \msun) results in stars that begin to burn hydrogen while still accreting.  The resulting radiation pressure should expel the accreting material preventing more massive stars from forming, yet stars much more massive do exist \citep{Larson1971,Zinnecker2007,Krumholz2009}.  At least two competing theories have been proposed to explain this paradox.  In the competitive accretion model, stars within a common gravitational potential accrete from a distributed gas reservoir.  The potential funnels gas down toward the cluster center, so the stars located closer to the center of the potential acrete at higher rates than isolated stars \citep{Bonnell1997,Bonnell2001,Bonnell2006}.  In the turbulent core accretion model, molecular clumps fragment into gaseous cores that subsequently collapse to make individual stars.  As gas accretes onto the protostar, gravitational energy is released that heats up the surrounding gas.  This increase in gas temperature raises the Jeans mass which suppresses fragmentation and results in fewer, but more massive, stars.  This monolithic collapse would dictate a correlation of the clump mass with the stellar mass, as the mass of the core determines the mass reservoir available to form the star \citep{Mckee2003,Krumholz2005,Krumholz2007}.  These divergent theories have led to increased interest in studying star-forming regions that occupy the transition from the low- to high-mass regimes.

\citet{Arvidsson2010} defined and introduced a small sample of 70 intermediate-mass star-forming regions (\imsfrs) --- regions producing stars up to but not exceeding $\sim$8 \msun.  IM SFRs populate the continuum linking the low- and high-mass star-forming regimes  where the onset of cluster formation may occur and the mechanisms that differentiate high-mass from low-mass star formation may become significant.  \imsfrs\ are sites of star formation distinct from regions of high-mass star formation.  They have typical luminosities of $\sim$$10^4$ \lsun\ and typical diameters of $\sim$1 pc, as judged from the sizes of the surrounding photodissociation regions (PDRs).  They are typically associated with molecular clumps of mass $\sim$$10^{3}$ \msun.  The molecular material associated with IM SFRs typically has an order of magnitude lower peak mass column density and clump mass compared to ultra-compact HII regions which are examples of high-mass star formation \citep{Arvidsson2010}.

We have begun a program to study the stellar properties and the local environments surrounding \imsfrs\ and identify a larger more complete sample of \imsfrs.  Our primary goal in this paper is to identify the stellar content and confirm the intermediate-mass nature of these candidate \imsfrs.  In Section 2 we define an all-sky sample of \imsfrs\ and categorize them based on their mid-IR morphological appearance.  In Section 3 we use the Two Micron All-Sky Survey (2MASS) Point Source Catalog \citep{Cutri2003}, new optical spectroscopy, and YSO SED fitting to investigate the stellar content of a sample of four \imsfrs.  In Section 4 we present the results of our stellar content analysis, and we summarize key results in Section 5.

\section{Sample Selection \label{sample.sec}}

\citet{Kerton2002} used \emph{Infrared Astronomical Satellite} (\emph{IRAS}) mid-infrared fluxes and \ion{H}{1} 21-cm emission maps to detect photodissociation regions (PDRs) associated with embedded intermediate-mass stars.  These were used to define candidate \imsfrs\ using a set of color selection criteria that specifies cool dust and large polycyclic aromatic hydrocarbon (PAH) contribution:  $-0.18 < \rm{log}(\rm{F}_{\nu}(25)/\rm{F}_{\nu}(12)) < 0.4, 0.96 < \rm{log}(\rm{F}_{\nu}(60)/\rm{F}_{\nu}(25)) < 1.56,$ and $0.29 < \rm{log}(\rm{F}_{\nu}(100)/\rm{F}_{\nu}(60)) < 0.69$.  For all candidate \imsfrs, the \emph{IRAS} flux density quality flag was required to be either high ($\rm{F}_{\mathrm{qual}}\,=\,3$) or moderate ($\rm{F}_{\mathrm{qual}}\,=\,2$).

\citet{Arvidsson2010} adopted the \citet{Kerton2002} criteria to define a sample totaling 984 candidate IM~SFRs and used $Spitzer$ Galactic Legacy Infrared MidPlane Survey Extraordinaire (GLIMPSE I) \citep{Benjamin2003} and MIPSGAL I \citep{Carey2005} legacy surveys to morphologically classify 70 inner-Galaxy targets that had complementary $^{13}$CO spectra from the Boston University Five College Radio Astronomy Observatory (BU-FCRAO) Galactic Ring Survey (GRS; \citealt{Jackson2006}).  They classified the candidate objects into three groups based on their mid-IR morphologies: ``blobs/shells,'' ``filamentary,'' and ``star-like.''  \citet{Arvidsson2010} found that the blobs/shells are star-forming regions characterized by circular or elliptical morphologies in the 8.0 and 24 \micron\ images.  The filamentary objects appear to be small knots along larger filamentary structures including regions where filaments intersect.  Star-like objects exhibit compact emission from one or more point sources in the mid-IR bands.  The \citet{Arvidsson2010} pilot study classified 49 (70.0\%) objects as blobs/shells, 11 (15.7\%) objects as filamentary, and 10 (14.3\%) as star-like.

\citet{Arvidsson2010} measured molecular column densities for 28 blobs/shells, of which 15 had distance determinations.  All 15 blob/shells with distance determinations, and thus masses, lie below the 1~g~cm$^{-2}$ minimum mass column density threshold for high-mass star formation posited by \citet{Krumholz2008}.  They found that these regions lack or have weak thermal radio continuum, limiting the most massive star(s) to early B or later, i.e., stars capable of exciting the PAHs, but not ionizing hydrogen.  \citet{Arvidsson2010} also found typical luminosities of $\sim$10$^4$ \lsun, where a luminosity of 10$^3$--10$^4$ corresponds roughly to a single $\sim$4--10 \msun\ young stellar object \citep{Robitaille2006}.  On the basis of these properties they concluded that these regions host, at most, intermediate-mass stars, though the stellar content was not directly observed.

\begin{figure}
\centering
\includegraphics[scale=0.85]{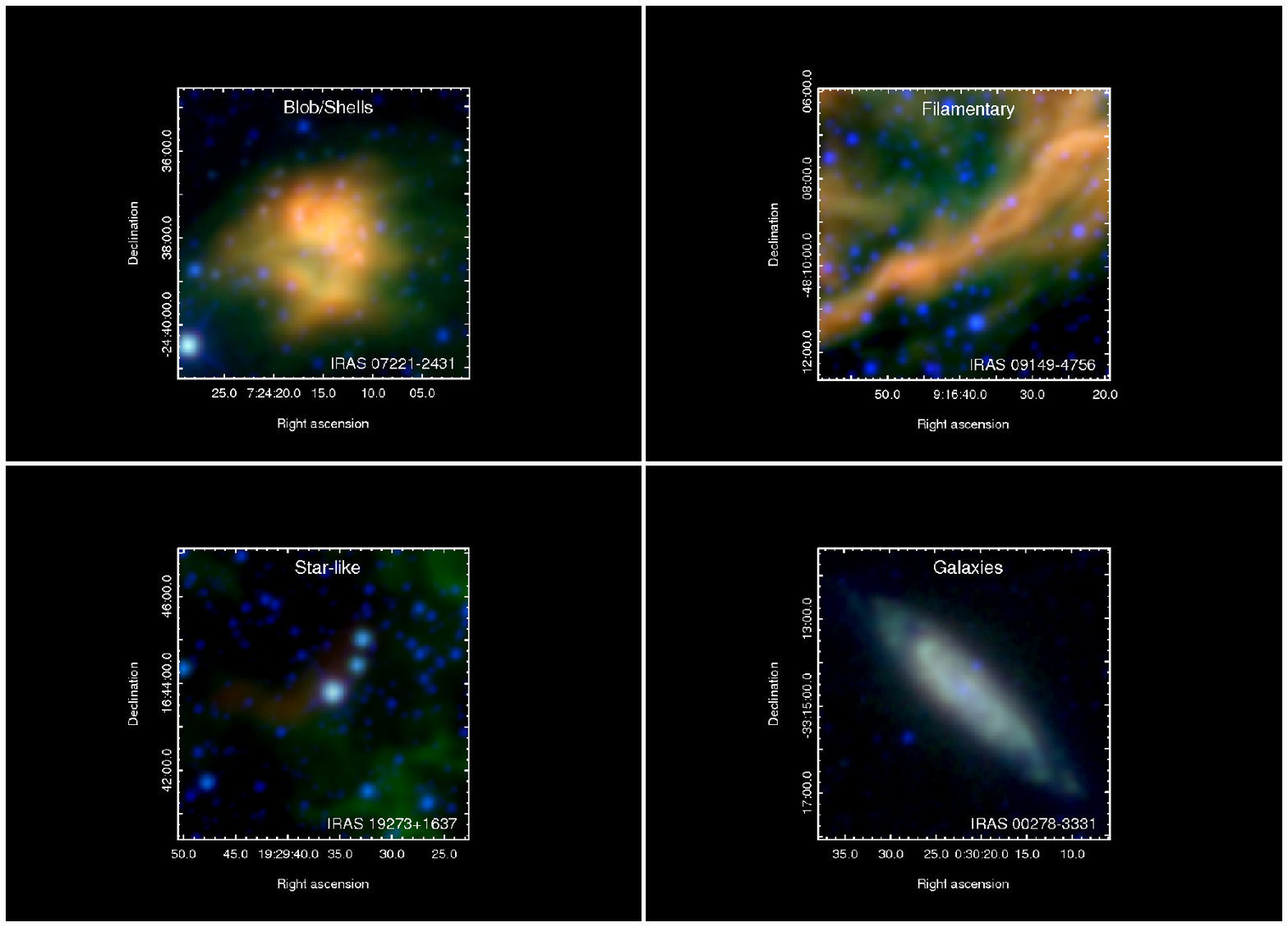}
\caption{A three-color rgb image ($WISE$ 22 \micron, 12 \micron, 3.4 \micron) showing a typical example of objects that fulfill the \emph{IRAS} color selection criteria for each morphological type:  blobs/shells, filamentary, star-like, and galaxies. \label{morphexample}}
\end{figure}

\begin{figure}
\centering
\includegraphics[scale=0.4]{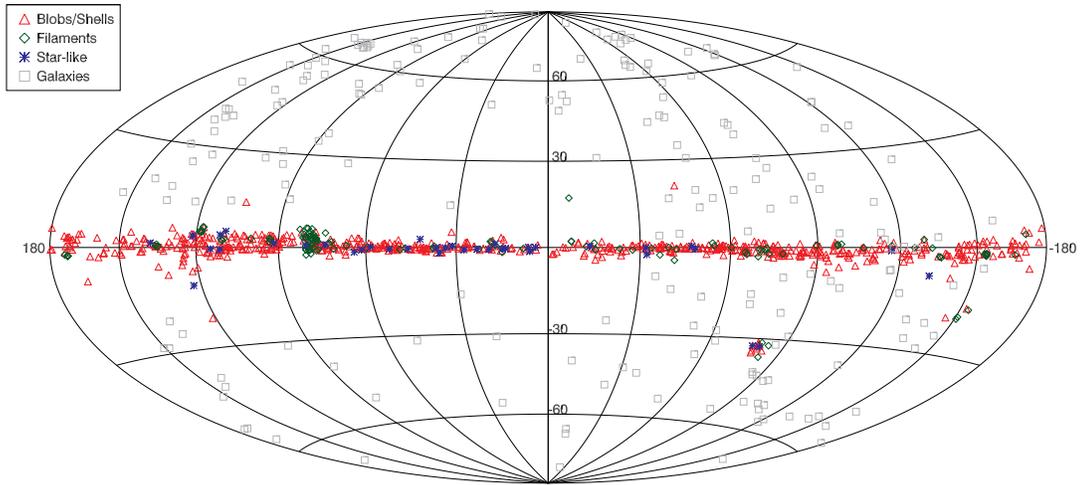}
\caption{Aitoff projection in Galactic coordinates showing the spatial distribution of 984 candidate \imsfrs.  The candidate IMSFRs are color-coded by morphological classification.  Red triangles correspond to blobs/shells, green diamonds to filamentary objects, blue stars to star-like objects, and grey squares to galaxies.  The blobs/shells, filamentary objects, and star-like objects are concentrated in the Galactic Plane with a small grouping in the Magellanic Clouds.  (A color version of this figure is available in the online journal.)  \label{allsky}}
\end{figure}

We used \emph{Wide-field Infrared Survey Explorer} (\emph{WISE}; \citealt{Wright2010}) 12 and 22 \micron\ images to extend the morphological classifications to the entire sample of 984 candidate \imsfrs.  We introduce a new category, ``galaxies'', to the ``blobs/shells,'' ``filamentary,'' and ``star-like'' scheme of \citet{Arvidsson2010}.  Figure~\ref{morphexample} is a three-color rgb image ($WISE$ 22 \micron, $WISE$ 12 \micron, $WISE$ 3.4 \micron) that shows  typical objects for each morphological category.  The blobs/shells are characterized by extended emission at 12 and 22 \micron, including isolated blobs, shells, and also enhancements at the edges of pillars or bright-rimmed clouds.  The filamentary objects appear to be long, narrow, low-surface-brightness whisps of emission at 12 and 22 \micron.  The star-like objects appear as isolated point sources (occasionally multiple) in all \emph{WISE} bands.  The galaxies exhibit extended emission that appears circular or lenticular in all \emph{WISE} bands, but they also have extended emission in near-infrared 2MASS and optical DSS images, rendering their morphology unmistakable.  Of the 984 candidate \imsfrs, 616 (62.6\%) are classified as blobs/shells, 128 (13.0\%) are classified as filamentary, 39 (4.0\%) are classified as star-like, and 201 (20.4\%) are classified as galaxies.  Table~\ref{imsfrsources.tbl} lists the \emph{IRAS} source name, coordinates, and morphological classification for all 984 objects in the sample.  Figure~\ref{allsky} shows the distribution of these regions projected on the sky in Galactic coordinates.  Red triangles correspond to blobs/shells, green diamonds to filamentary, blue stars to star-like, and grey squares to galaxies.  Figure~\ref{allsky} illustrates that galaxies dominate the sample at high Galactic latitudes while blobs/shells, filamentary structures, and star-like objects are heavily concentrated toward the Galactic Plane.  There is a conspicous overdensity of filamentary objects that constitutes nearly one-third of all filamentary objects, 41 of the 128, around $\ell$ = 80\deg\ corresponding to the Cygnus-X region.  This overdensity may be a result of the massive star content \citep{odenwald1993} coupled with the relative proximity of the region at $\sim$1.2--1.4 kpc \citep{Kobulnicky2013AAS}.  There is also an overdensity of nine objects classified as blobs/shells around $\ell$ = 280\deg, $b$ = -32\deg, at the location of the Large Magellanic Cloud.  Another object classified with the blobs/shells (IRAS~00402+4120) is located in M31. 

Taken together, the distribution of the morphological classes on the sky is
consistent with the blobs/shells and filaments being star-forming regions (or
affiliated with star-forming regions) in the Galactic Plane. The uniform
distribution of galaxies is consistent with their classification as extra-Galactic
sources.  The nature of the star-like objects is not known.  Some of these are
likely to be very compact star-forming regions, perhaps akin to the Ultra-Compact
Embedded Clusters (UCECs) identified by \citet{Alexander2012}.  Some of these may be
post- or pre- main sequence objects with circumstellar disks, such as
IRAS~00487+5118 (\#22 in Table~1) a Vega-type excess source \citep{Dent2005}.  Higher
resolution mid-IR imaging and/or infrared spectroscopy would help discern the nature
of these sources.

\begin{figure}
\centering
\includegraphics[scale=0.75]{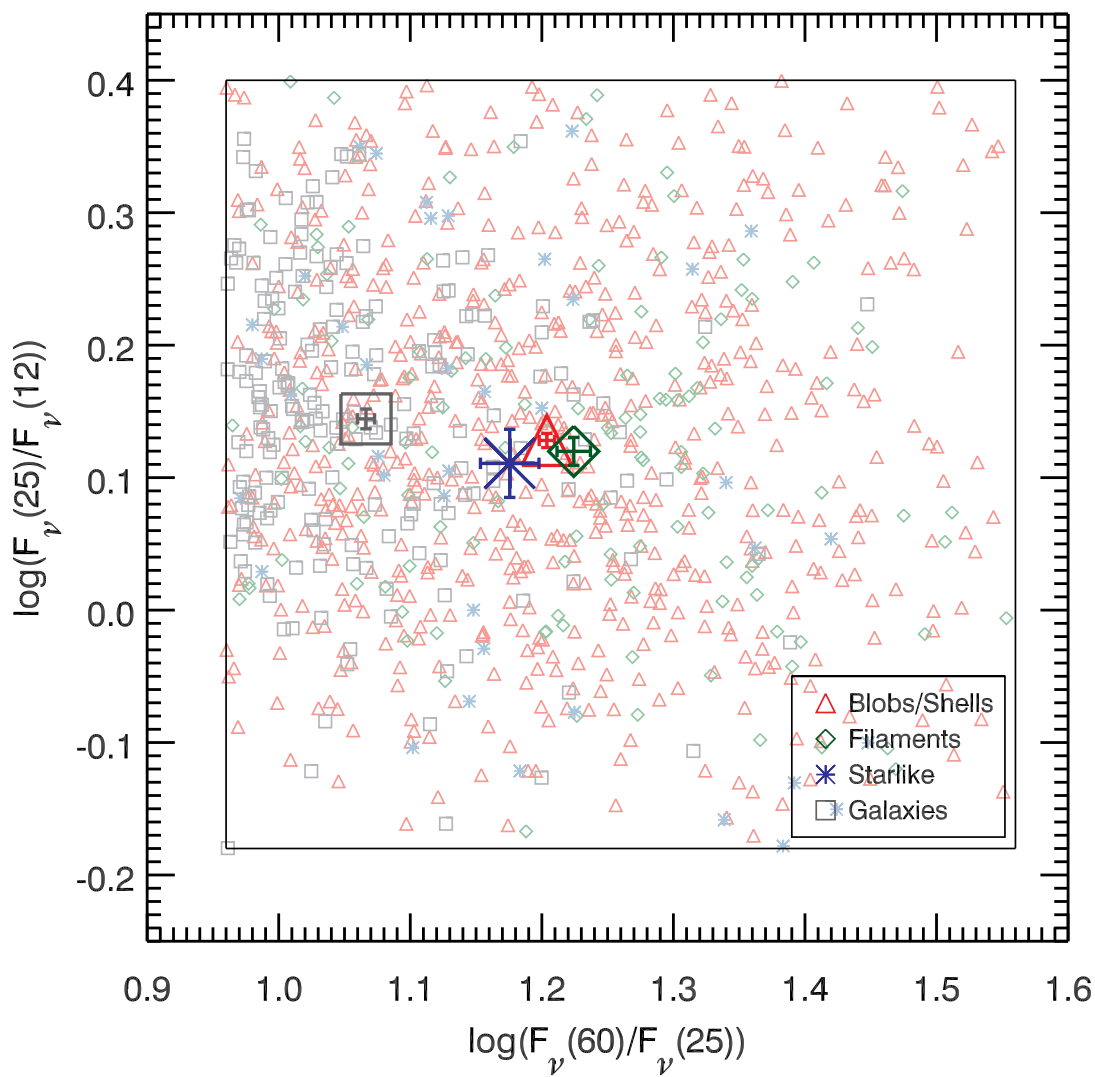}
\caption{\emph{IRAS} color-color diagram of the 984 candidate \imsfrs.  The box indicates the boundaries of the \emph{IRAS} color selection for the sample.  Red triangles correspond to blobs/shells, green diamonds to filamentary objects, blue stars to star-like objects, and grey squares to galaxies.  The larger, darker colored symbols, represent the average colors for each morphological type with error bars indicating the error of the mean.  (A color version of this figure is available in the online journal.)  \label{irasccd}}
\end{figure}

Figure~\ref{irasccd} plots the 984 candidate IMSFRs in a $\rm{log}(f_{\nu}(25)/f_{\nu}(12))$ vs. $\rm{log}(f_{\nu}(60)/f_{\nu}(25))$ \emph{IRAS} color-color diagram.  The box indicates the \citet{Kerton2002} color selection domain, and the symbols denote candidate regions by morphological type as in Figure~\ref{allsky}.  The larger symbols with error bars indicate the mean color and error of the mean for each morphological type.  The $\rm{log}(f_{\nu}(60)/f_{\nu}(25))$ color is a measure of dust temperature while the $\rm{log}(f_{\nu}(25)/f_{\nu}(12))$ color is sensitive to the PAH contribution.  While the galaxies are more concentrated toward the higher dust temperatures at the left of the diagram, the mean colors of the blobs/shells, filamentary, and star-like objects are indistinguishable.   The density of points drops markedly toward the right edge (low dust temp) and bottom edge (high PAH-to-dust ratio) of the plot, suggesting that a more generous color selection box would not yield significantly more objects.  Given the combination of our color criteria and follow-up morphological classification, it is likely that our list of 984 candidate objects contains a reasonably complete sample of intermediate-mass star forming regions, at least to the limits imposed by \emph{IRAS} sensitivity and beamsize (1.5\arcmin$\times$4.7\arcmin\ at 60 \micron).  The sample list could be expanded by a more generous color selection box toward higher dust temperatures, but this would lead to larger contamination by galaxies and require additional visual examination.  Expanding the color selection toward smaller PAH-to-dust content would pick up additional star forming regions, but many of these would be increasingly dominated by massive, ionizing stars (e.g., see \citealt{vanderveen1988,Walker1989,Zhang2004}).

\section{A Pilot Survey of Stellar Content in Four Intermediate-Mass Star-Forming Regions \label{imregions.sec}}

While \citet{Arvidsson2010} studied the gaseous content of a sample of inner-Galaxy \imsfrs, the stellar content of these regions remains unexplored.  We have selected four outer-Galaxy intermediate-mass regions classified as blobs/shells and having four or more optically visible stars for an initial study of the stellar content of \imsfrs.  The regions in this initial study are isolated and at high Galactic latitudes where there is expected to be less extinction and little confusion with other Galactic star forming regions.  Table~\ref{sources.tbl} lists the \emph{IRAS} names, central positions, and \emph{IRAS} fluxes of these sources.  Figures~\ref{4reg1} and \ref{4reg2}  show three-color rgb images (\emph{WISE} 22 \micron, 12 \micron, 3.4 \micron) of the four \imsfrs.  Cyan crosses mark the locations of optically visible stars selected for spectroscopic study (discussed in Section~\ref{spec.sec} below).  White circles indicate 2--3.5\arcmin\ diameter regions used to search for possible stellar clusters using photometry from the Two Micron All Sky Survey (2MASS) Point Source Catalog \citep{Cutri2003}.  Red circles indicate candidate YSOs.  Black diamonds indicate maser sources, and white diamonds indicate sub-mm sources.

\subsection{Infrared Photometry \label{irphot.sec}}

\begin{figure}
\centering
\includegraphics[scale=0.9]{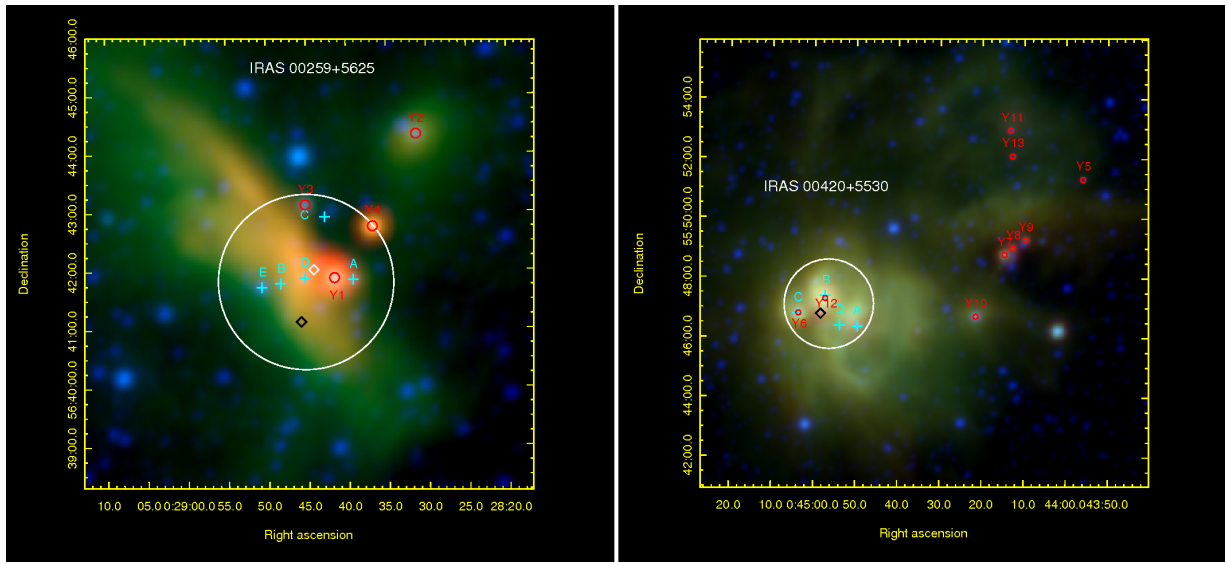}
\caption{Three-color rgb image ($WISE$ 22 \micron, 12 \micron, 3.4 \micron) of IRAS~00259+5625 (left) and IRAS~00420+5530 (right).  Cyan crosses indicate individual stars selected for optical spectroscopy, and white circles indicate the regions regions used to search for possible stellar clusters.  White diamonds mark the positions of known sub-mm sources and black diamonds mark the positions of known masers. \label{4reg1}}
\end{figure}

\begin{figure}
\centering
\includegraphics[scale=0.9]{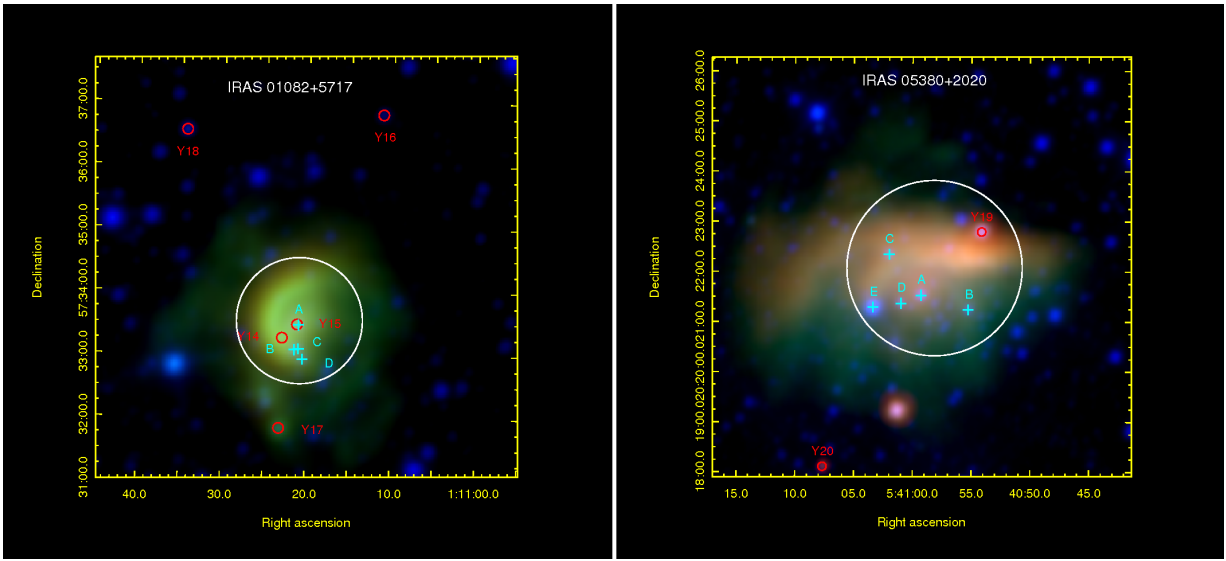}
\caption{As Figure~\ref{4reg1} for IRAS~01082+5717 (left) and IRAS~05380+2020 (right). \label{4reg2}}
\end{figure}

\begin{figure}
\centering
\includegraphics[scale=0.6]{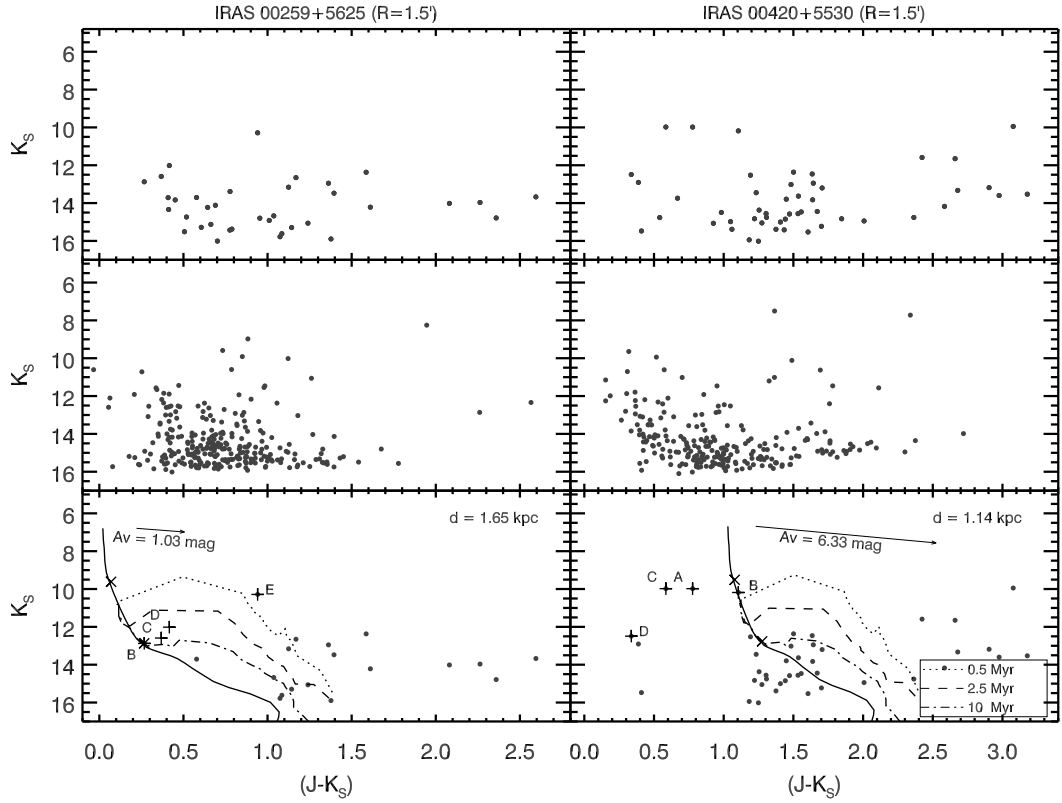}
\caption{CMDs of IRAS~00259+5625 (left column) and IRAS~00420+5530 (right column) for the defined cluster region (top row), comparison field annulus (middle row), and the cleaned cluster region (bottom row).  The lower panel shows a reddened Padova ZAMS isochrone (solid curve) with upper and lower X's indicating B0V and A0V locations, respectively.  The cleaned cluster also shows reddened Siess 0.5 Myr (dotted line), 2.5 Myr (dashed curve), and 10 Myr (dotted-dashed curve) pre-main-sequence tracks.  Pluses indicate stars selected for spectroscopy.  The adopted distance and extinction are given in the lower panels. Arrows indicate the direction and magnitude of interstellar reddening, as labeled. \label{cmd1}}
\end{figure}

\begin{figure}
\centering
\includegraphics[scale=0.6]{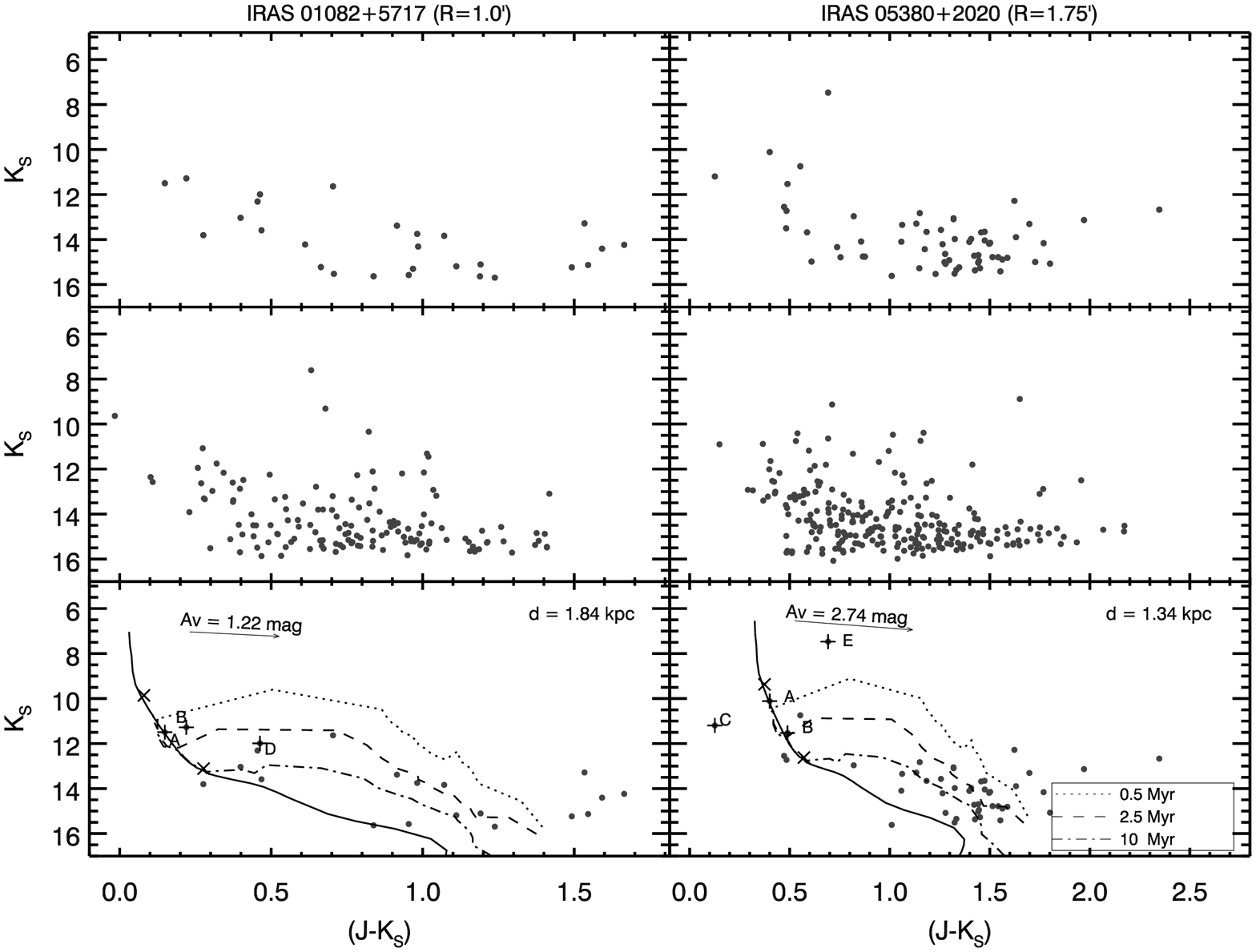}
\caption{CMDs, as in Figure~\ref{cmd1}, for IRAS~01082+5717 (left column) and IRAS~05380+2020 (right column). \label{cmd2}}
\end{figure}

We used the 2MASS Point Source Catalog to generate \jhks\ color-magnitude diagrams (CMDs) and color-color diagrams (CCDs) of each region. Figures~\ref{cmd1} and \ref{cmd2} (top row) plot the raw \ks\ vs. $J-K_{S}$ diagram for each region enclosed within the circles shown in Figures~\ref{4reg1} and \ref{4reg2}.  As field stars can outnumber the associated main-sequence and pre-main sequence stars in a region, they must be removed statistically from the CMD to recover the likely cluster stars.  We use the cluster cleaning tool of \citet{Alexander2013} which follows the prescription of \citet{Maia2010} to evaluate and remove the field star contamination.  The middle row of panels in Figure~\ref{cmd1} and \ref{cmd2} show the \ks\ vs. $J-K_{S}$ diagrams for the field region.  The field region is an annulus surrounding the cluster aperture that covers eight times the area of the cluster aperture.  The field star estimation tool compares a target region to a field region using a 2MASS \jhks\ 3-dimensional color-color-magnitude diagrams (CCMD).  After investigating different color and magnitude combinations for the CCMDs, we used the $J$, $J - H$, and $J - K_{S}$ magnitudes, as in \citet{Maia2010} for field subtraction.  The lower row of panels in Figure~\ref{cmd1} and \ref{cmd2} show the field-subtracted \ks\ vs. $J-K_{S}$ cluster CMDs.  The solid line is a Padova \citep{Marigo2008,Girardi2010} zero-age main sequence that has been placed at the estimated distance and reddening (discussed in Section \ref{results.sec} below).  The dotted, dashed, and dot-dashed lines indicate \citet{Siess2000} 0.5, 2.5, and 10.0 Myr pre-main-sequence isochrones respectively.  Stars with optical spectra are indicated by pluses and labeled by spectral type.

\begin{figure}
\centering
\includegraphics[scale=0.6]{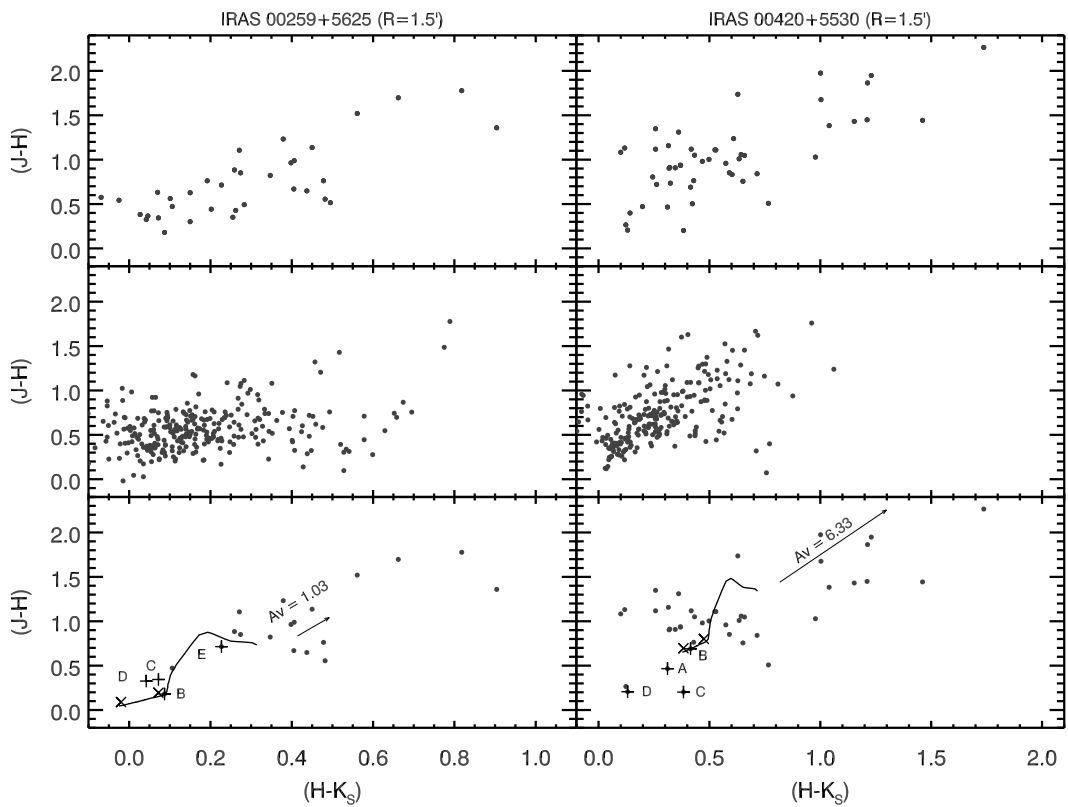}
\caption{CCDs of IRAS~00259+5625 (left column) and IRAS~00420+5530 (right column) for the defined cluster region (top row), comparison field annulus (middle row), and the cleaned cluster region (bottom row).  The solid curve in the lower panel shows a Padova ZAMS isochrone with upper and lower X's indicating the locations of B0V and A0V stars, respectively.  Pluses indicate stars selected for spectroscopy.  The extinction is given for each cleaned cluster, and arrows indicate the direction of the reddening vectors. \label{ccd1}}
\end{figure}

\begin{figure}
\centering
\includegraphics[scale=0.6]{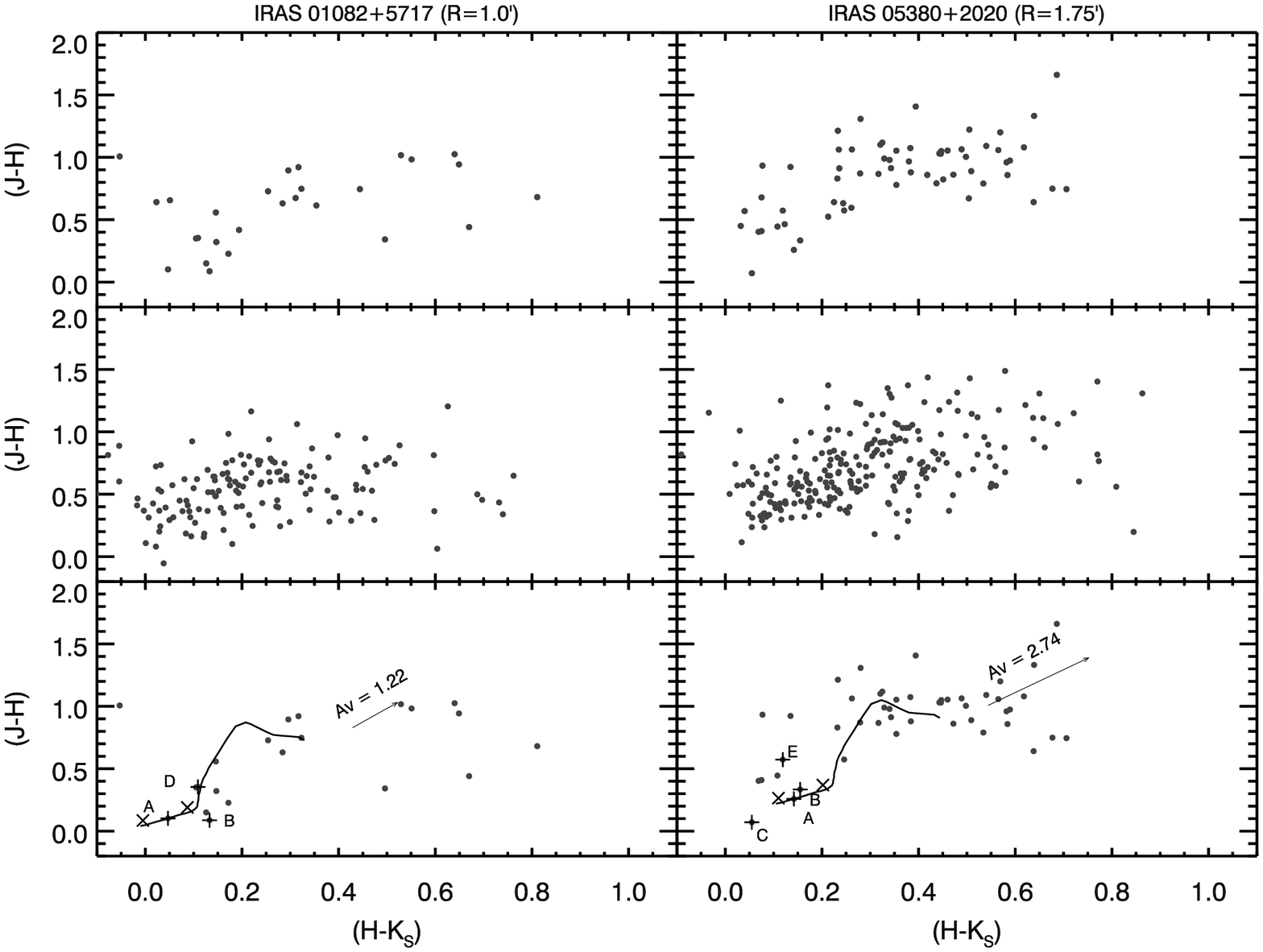}
\caption{CCDs, as in Figure~\ref{ccd1}, for IRAS~01082+5717 (left column) and IRAS~05380+2020 (right column). \label{ccd2}}
\end{figure}

Figure~\ref{ccd1} and \ref{ccd2} show the $J-H$ vs. $H-K_{S}$ CCDs cleaned in the same manner as Figure~\ref{cmd1}. The top row plots the raw $J-H$ vs. $H-K_{S}$ CCDs for each region enclosed within the circles shown in Figures~\ref{4reg1} and \ref{4reg2}.  The middle row of panels plots $J-H$ vs. $H-K_{S}$ for the field region.  The lower row of panels in Figure~\ref{ccd1} and \ref{ccd2} show the field-subtracted $J-H$ vs. $H-K_{S}$ cluster CCDs.  The solid line is a Padova \citep{Marigo2008,Girardi2010} ZAMS that has been placed at the estimated reddening (discussed in Section \ref{results.sec} below).  Stars having optical spectroscopy are indicated by pluses and labeled by spectral type.

We interpret these CMDs and CCDs in Section \ref{results.sec} to look for compact stellar populations of similar reddening and distance.

\subsection{Optical Spectroscopy \label{spec.sec}}

We used the CMDs and CCDs described above in conjunction with optical catalogs to select sources within the four candidate \imsfrs\ having bright $K_{S}$-band magnitudes and $V\lesssim$16.  This $K_{S}$-band magnitude criterion includes the stars that are likely to be the most luminous and, therefore, the most massive cluster members.  We obtained optical spectra for eighteen sources using the LongSlit spectrograph on the Wyoming Infrared Observatory (WIRO) 2.3 m telescope in 2011 December and 2012 September.  Each object lies within the central 1--2\arcmin\ of the four candidate \imsfrs.  The data for nine of the eighteen objects were obtained with a 600 l mm$^{-1}$ grating giving a spectral coverage of 4000--7000 \ang\ at a typical reciprocal dispersion of $\sim$1.5 \ang\ pixel$^{-1}$ across the chip.  The remaining nine spectroscopic targets were observed with a 2000 l mm$^{-1}$ grating giving a spectral coverage of 5400--6800 \ang\ at a typical reciprocal dispersion of $\sim$0.72 \ang\ pixel$^{-1}$.  The typical signal to noise ratio at 5500 \ang\ was greater than 40 with several stars exceeding 200.  The data were reduced in the standard manner using dome lamp flat-fielding, bias frames, and CuAr comparision lamp spectra for wavelength calibration to a typical precision of 0.015 \ang\ rms for the 2000 l mm$^{-1}$ grating and 0.05 \ang\ rms for 600 l mm$^{-1}$. 

\subsection{Spectral Classification}

Extinction by dust along the line of sight produces low signal-to-noise in the typical temperature diagnostic \ion{He}{1} $\lambda$4471 \ang\ and \ion{Mg}{2} $\lambda$4481 \ang\ lines, making classifications difficult for B- and early A-type stars.  In our spectra the available lines that are useful for spectral classification of these stars are the \ion{He}{2} $\lambda$5411 \ang, \ion{He}{1} $\lambda$5876 \ang, and H$\alpha$ $\lambda$6563 \ang\ lines.  The absence of the \ion{He}{2} $\lambda$5411 \ang\ line in all of our spectra limits the spectral type to B0.5 or later.  The absence of the \ion{He}{1} $\lambda$5876 \ang\ line in some of the targets limits the spectral type to later than about A0.  The EW of the H$\alpha$ line is at a maximum at a spectral type of about A1, but is double-valued, decreasing in strength for stars both earlier and later.  

Spectral classification was done by comparision with the spectral atlas of
\citet{Jacoby1984} and by comparison to the spectral indices of
\citet{Hernandez2004} and \citet{Kobulnicky2012}.  For B-stars this included
comparisions of the equivalent widths of H$\alpha$  and \ion{He}{1} $\lambda$5876
\ang.  For A- and F-stars this included comparisions of the equivalent width of
H$\alpha$.  For G-stars this included comparisions of the equivalent width of
H$\alpha$ and  the \ion{Mg}{1} triplet at
$\lambda\lambda5167,5172,5183$ \ang.  For K-stars this included comparisions of the
equivalent width of H$\alpha$ and visual comparisions with the MgH band at
$\lambda5000$--$5200$ \ang.  Spectral classifications of B-, A- and F-type stars are
typically certain to $\pm$1 spectral sub-type for a typical signal to noise ratio of
40 using this method. \ion{He}{1} and H$\alpha$ lack the sensitivity to
differentiate B0.5--B3 stars, so the two stars that appear to fall in this range
have been classified as B2. Owing to interstellar extinction and limited wavelength coverage, no stars have the combination of good signal-to-noise in the blue part of the optical range needed for luminosity classification; in the absence of constraining data, we assume that all stars are luminosity class V, but we caution that this may not always be the case.

Table~\ref{2mass.tbl} lists the relevant parameters for determining the spectral types and spectrophotometric distances for each star.  Column 1 lists the \emph{IRAS} region name and the corresponding alphabetic identifier in Figures~\ref{4reg1} and \ref{4reg2} for each star with optical spectra.  Column 2 lists the 2MASS ID.  Columns 3--5 list the $J$-, $H$-, and \ks- band magnitudes given by the 2MASS Point Source Catalog.  Columns 6--9 list the equivalent widths and equivalent width uncertainties for H$\alpha$ and \ion{He}{1}, the primary spectral lines used to determine spectral types.  Column 10 lists the estimated spectral type.  Column 11 shows the extinction determined from the 2MASS magnitudes and the spectral type including upper and lower limits assuming an uncertainty of $\pm$1 sub-type for B-, A-, and F- type stars and $\pm$3 sub-types for G-type and later stars.  Column 12, D$_{spec}$, is the spectrophotometric distance determined from the 2MASS magnitudes and the spectral type, including upper and lower limits, where the upper/lower limits result from the random uncertainty in spectral type and the second {\it upper-only} limit accounts for systematic uncertainty resulting from the unknown binarity status of each star; if any of these stars are close binaries with equal-luminosity companions  \citep[common among massive stars;][]{Kiminki2012}, the distances could be underestimated by as much as $\sqrt{2}$.  Column 13, D$_{lit}$, lists distances found in literature for each region.  Estimates of the distance to each region were done by using 2MASS $H-K_{S}$ color excesses to derive \av\ and \ak\ using a \citet{Cardelli1989} exctinction curve.  A distance modulus was calculated with \ak, the 2MASS \ks-band magnitude, and the intrinsic \ks-band magnitude from a Padova zero age main-sequence (ZAMS) isochrone, assuming the observed stars are dwarfs (luminosity class V).  The uncertainties on the distance estimates are dominated by the uncertainties on our spectral classifications and the (unknown) binary status.  

\subsection{Young Stellar Objects}
The young stellar objects (YSOs) near each region were identified by generating spectral energy distributions (SEDs) using 2MASS \jhks\ near-IR and \emph{WISE} 3.4, 4.5, 12, and 22 \micron\ mid-IR data with source cross-band matching done from the \emph{WISE} Point Source Catalog which includes 2MASS.  These sources were fit using the SED fitting tool of \citet{Robitaille2007} in a similar manner to \citet{Alexander2013}.

All \emph{WISE} point sources within 10\arcmin\ of the \imsfr\ were selected for fitting.  We required a minimum of three photometric data points per target and removed objects that did not meet this criterion.  We first fit sources to \citet{Kurucz1993} stellar models where we allowed the extinction parameter to vary from \av\ = 0 to 40 magnitudes.  Sources with $\chi^{2}/\mathrm{n}_{data} \leqslant 2$ were deemed to be well fit by a reddened stellar photosphere and removed.  To remove sources with bad fits due to high mid-IR sky backgrounds, the remaining sources were visually inspected at \emph{WISE} bands W3 and W4 and removed if they did not appear to be point sources.  The SED fitting tool was then used to fit the remaining sources to a grid of 200,000 YSO model SEDs \citep{Robitaille2006}.  Table~\ref{ysoinput.tbl} lists the input parameters for this fit.  Column 1 indicates which IRAS target field contains the YSO.  Column 2 lists the assigned YSO ID.  Column 3 gives the \emph{WISE} ID.  Columns 4--6 indicate the 2MASS PSC magnitudes.  Columns 7--10 list the \emph{WISE} PSC magnitudes.  We chose model aperture sizes of 3\arcsec\ for 2MASS, 6\arcsec\ for \emph{WISE} bands W1, W2, and W3, and 12\arcsec\ for \emph{WISE} band W4 for all regions, while the extinction parameter varied from \av\ = 0 to 35 magnitudes for all regions.  We allowed the distance parameter to vary from 1.5--2.5 kpc for IRAS~00259+5625, 2.0--2.5 kpc for IRAS~00420+5530 (best estimate maser distance), 1.0--2.0 kpc for IRAS~01082+5717, and 0.7--1.5 kpc for IRAS~05380+2020.  We selected models where the fits were limited to $(\chi^{2}-\chi^{2}_{best})/\mathrm{n}_{data} < 2$ and used a weighted average to estimate YSO properties in the manner of \citet{Alexander2013}.

Table~\ref{newysoresults.tbl} compiles the results of the YSO fitting.  This table lists the number of models well fit to each YSO SED and the $\chi^{2}/\mathrm{n}_{data}$ as a measure of the goodness of fit.  It also lists best fit and average values for the extinction, stellar mass, disk mass, and disk accretion rate, in addition to the range of stellar mass fit by the models and the total YSO luminosity.  These parameters were used to determine the stage of the YSO.  \citep{Robitaille2006} assigns stage 0/I for heavily embedded sources with envelope accretion rates $\dot{\mathrm{M}}/\mathrm{M}_{*}$ $>$ $10^{-6}$ $\mathrm{yr}^{-1}$.  Stage II YSOs have $\dot{\mathrm{M}}/\mathrm{M}_{*}$ $<$ $10^{-6}$ $\mathrm{yr}^{-1}$ and $\mathrm{M}_{disk}/\mathrm{M}_{*}$ $>$ $10^{-6}$.  They have little or no mass left in the envelope, but are young enough to still have a protostellar disk.  Stage III YSOs $\dot{\mathrm{M}}/\mathrm{M}_{*}$ $<$ $10^{-6}$ $\mathrm{yr}^{-1}$ and $\mathrm{M}_{disk}/\mathrm{M}_{*}$ $<$ $10^{-6}$.  They have little or no mass left in either an envelope or a protostellar disk.  Sources with fits that do not have a total weight (probability) greater than 0.67 for any single stage are assigned as ambiguous (amb).

\section{The Stellar Content of IM~SFRs \label{results.sec}}

\subsection{IRAS~00259+5625 \label{00259.sec}}
IRAS~00259+5625 was classified as an \imsfr\ in the Bok globule CB3 by \citet{Codella1999} due to its IR bolometric luminosity of 930 \lsun\ \citep{Launhardt1997}.  \citet{Launhardt1997} found a kinematic distance of 2.5 kpc assuming that IRAS~00259+5625 was associated with the near side of the Perseus arm.  \citet{Yun1994a} obtained $^{12}$CO velocity distance of 2.1 kpc using the \citet{Clemens1985} rotation curve.  There is a compact submillimeter source likely consisting of an aggregate of multiple Class 0 sources 15\arcsec\ west and 6\arcsec\ south of the IRAS~00259+5625 source position \citep{Huard2000} indicated by the white diamond in the left panel of \ref{4reg1}.  The black diamond in the left panel of \ref{4reg1} indicates an $\mathrm{H}_{2}\mathrm{O}$ maser also present in the region \citep{Scappini1991}.

\begin{figure}
\centering
\includegraphics[scale=0.65,angle=90]{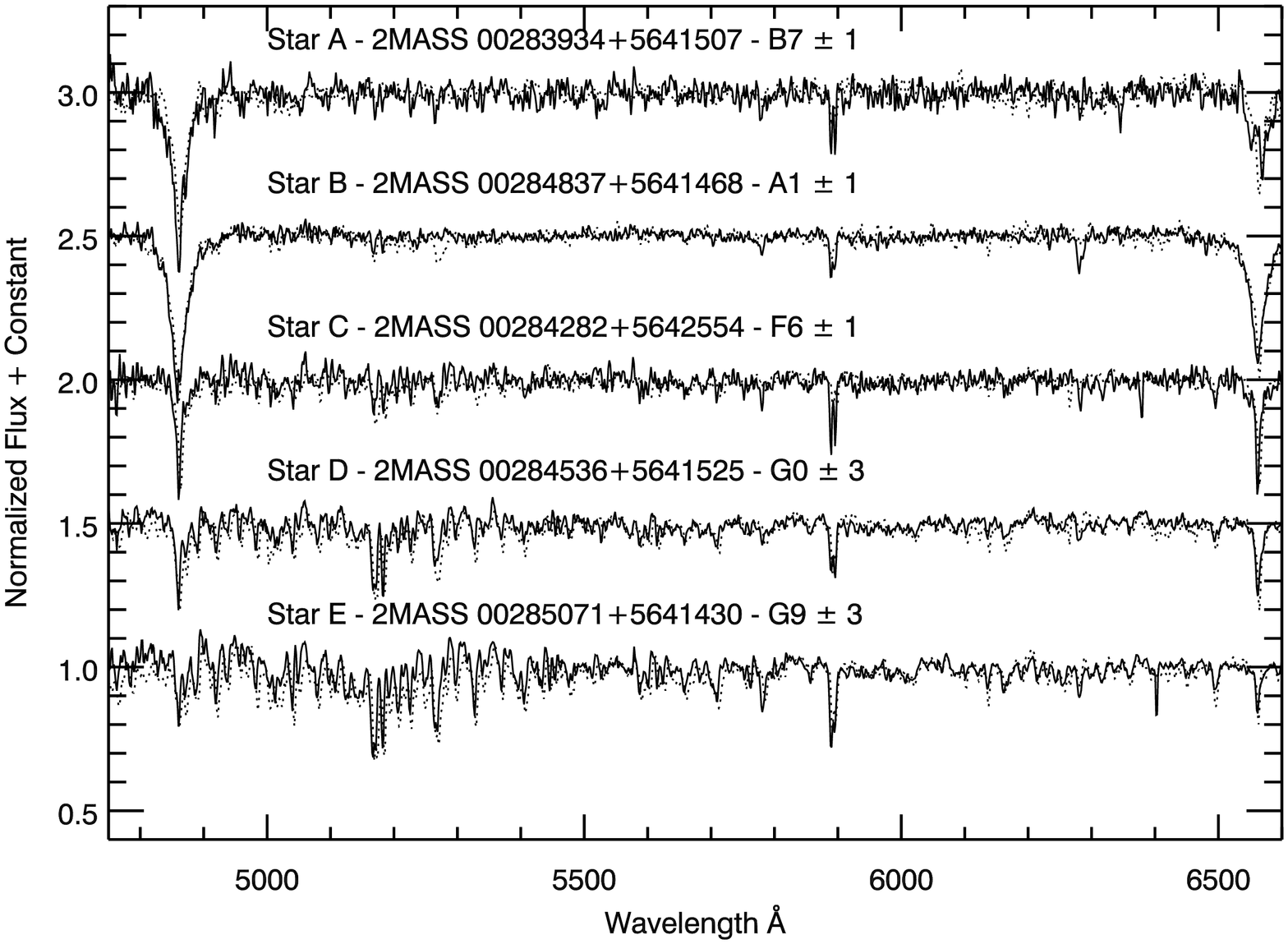}
\caption{Continuum-normalized spectra (solid black lines) of the five stars selected for optical spectroscopy from IRAS~00259+5625.  The spectra are overplotted with red dotted lines of comparable spectral type from the \citet{Jacoby1984} spectral atlas. \label{spec00259}}
\end{figure}

Figure~\ref{spec00259} shows our optical spectra of the five stellar targets.  Black solid lines denote our spectra, while the dotted lines show the best comparison spectra from the \citet{Jacoby1984} spectral atlas.  Star A has a weak \ion{He}{1} $\lambda$5876 \ang\ absorption feature ($EW$=0.29$\pm$0.13 \AA) and a strong H$\alpha$  absorption line ($EW$=7.99$\pm$0.57 \AA), indicative of a mid-B star (B7).  Star B has a spectrum with strong H$\alpha$ absorption ($EW$=14.95$\pm$0.17 \AA) indicative of an early A star (A1).  Star C has strong, narrow H$\alpha$ absorption (EW=4.53$\pm$0.19 \AA) indicative of an mid-F star (F6).  Star D has weak, narrow H$\alpha$ absorption and the presence of the \ion{Mg}{1} triplet at $\lambda\lambda5167,5172,5183$ \ang\ indicative of an early G star (G0).  Star E has a weaker, narrow H$\alpha$ absorption and an increasing presence of the \ion{Mg}{1} triplet at $\lambda\lambda5167,5172,5183$ \ang\ indicative of a late-G star (G9).  We used the spectrum of Star B and its 2MASS magnitudes and colors to derive a spectrophotometric distance of 1.65$^{+0.04+0.68}_{-0.08}$ kpc with an extinction of \av\ = 1.03$^{+0.04}_{-0.03}$ mag for the region.

IRAS~00259+5625 contains four candidate YSOs in the vincinity of the \imsfr.  These include three intermediate-mass YSOs (Y1, Y2, and Y3) at 4.0, 2.8, and 3.0 \msun\ and one lower-mass YSOs (Y4) at 0.8 \msun.  Of the three intermediate-mass YSOs, two are Stage II YSOs (Y2 and Y3), while the Y1 fit is ambiguous.  The lower-mass YSO, Y4, is Stage I.  The total luminosity contribution of these YSOs is 2.4$\times10^2$ \lsun.  Using the \citet{Sanders1996} definition for infrared luminosity, we use the \emph{IRAS} fluxes to derive an infrared luminosity of 3.1$\times10^2$ \lsun\ for the region, which is consistant with a total luminosity from the most massive YSOs identified and tabulated in Table~\ref{newysoresults.tbl}.  

The cleaned CMD of IRAS~00259+5625 in the left panel of Figure~\ref{cmd1} is rather sparse, showing just a few sources near the nominal main sequence at a distance of 1.65 kpc and reddening of \av\ = 1.03 mag.  Star B, found to be an early A star from our optical spectroscopy, is labeled and lies near the nominal A0V star (lower X) along the main sequence.  Star A ($\sim$B7) is not drawn because the 2MASS Catalog lists only upper limits for its J and H photometry as the source is detected, but not resolved from a nearby companion, in these bands.  Stars C and D (mid- to late-F) lie just to the right of the main sequence along the 10 Myr isochrone near where it joins the main sequence. These appear slightly too bright to be mid-F stars at the adopted cluster distance and reddening; they may be foreground objects at d $\simeq$ 0.6 kpc (see Table~\ref{2mass.tbl}), or they could be binary/multiple systems having magnitudes brighter than those of single stars of the equivalent type.  It is unlikely that they are overluminous pre-main-sequence objects because our optical spectra do not show the H$\alpha$ emission signatures of active accretion.  Star E, found to be a late-G type, lies well to the upper right of the main sequence (brighter and redder), consistent with it possibly being a field giant. A group of about eight stars lie along the 2.5 and 10 Myr isochrones, most of them near $J-K_S=1.2$, at the locations where late-stage pre-main-sequence stars would be expected. Another sequence of about eight stars stretch redward from pre-main-sequence tracks, consistent with them being either enshrouded pre-main-sequence stars or more reddened background objects.  Of the four YSOs in this region, three lie within the marked cluster aperture (Y1, Y3, and Y4), but only Y4 appears on Figure~\ref{cmd1} near $K=13.67$, $J-K_S=2.6$.  The other two lie off the plot because of their extreme colors ($J-K_S\sim4.6$).  The color-color diagram for IRAS~00259+5625 (Figure~\ref{ccd1}) reveals a loose group of stars near $H-K_S=0.4$, $J-H=1.0$, with a few stretching along the reddening vector to more extreme colors.  This grouping of stars may consist of moderately reddened (\av\ $\sim$4) intermediate-mass stars, lightly reddened low-mass stars, or some combination thereof.  Taken together, the cleaned CMD, CCD, and the spectra of IRAS~00259+5625 are consistent with a poor cluster of pre-main-sequence stars in the vincinity of the mid-B, early A, and late-F stars that appear to be the most massive constituents of this region.  The limited depth of the 2MASS photometry precludes a more definitive characterization of the possible stellar cluster in IRAS~00259+5625, though using the \citet{Weidner2013} most massive star -- cluster mass relation, we can estimate a total cluster mass of $\sim$20 \msun\ in $\sim$65 stars above 0.1 \msun\ after integrating a typical \citet{Kroupa2001} IMF.

\subsection{IRAS~00420+5530 \label{00420.sec}}

\citet{Kumar2006} investigated the IRAS~00420+5530 region and found it to contain a 380 \msun\ cluster.  \citet{Molinari2002} used millimeter data to determine that this region had a bolometric luminosity of $L=12,400$ \lsun\ at a kinematic distance of 5.0 kpc.  Based on its radio free-free flux, they also determined that a $\sim$$3000$ \lsun\ B2 ZAMS star was present.  \citet{Zhang2005} found a kinematic distance of 7.72 kpc.  \citet{Moellenbrock2009} obtained a distance of 2.17 kpc via trigonometric parallax of an $\mathrm{H}_{2}\mathrm{O}$ maser.  The maser is indicated by the black diamond in the right panel of Figure~\ref{4reg1}.  \citet{Koenig2012} noted a small cluster of YSOs associated with this region and considered it to be distinct from the nearby massive star-forming region NGC 281, for which \citet{Sato2008} obtained an $\mathrm{H}_{2}\mathrm{O}$ maser trigonometric parallax distance of 2.82 kpc.  \citet{Rygl2010} suggested that both IRAS~00420+5530 and NGC~281 may be on the edge of an expanding superbubble.

\begin{figure}
\centering
\includegraphics[scale=0.65,angle=90]{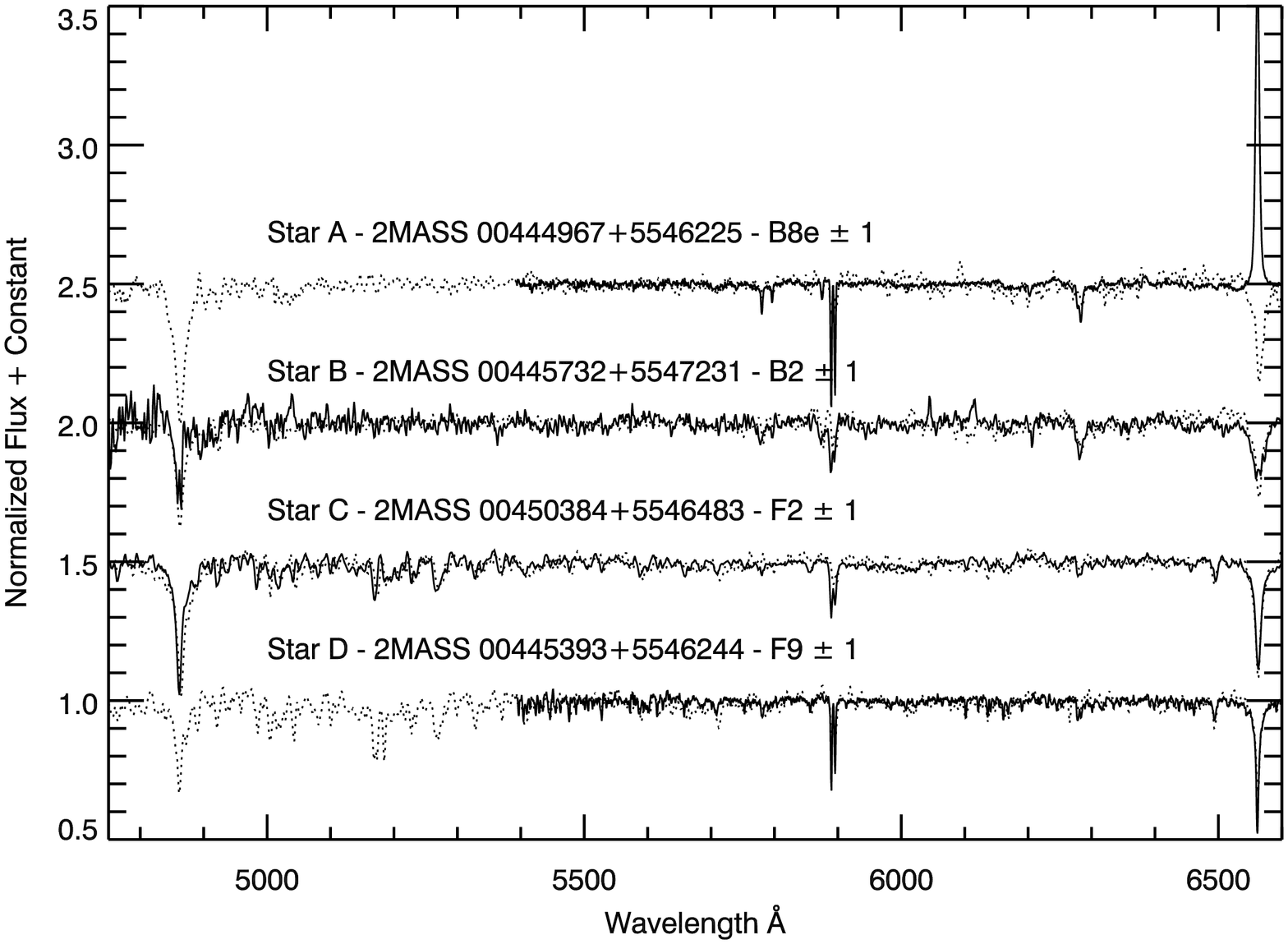}
\caption{Continuum-normalized spectra (solid black lines) of the five stars selected for optical spectroscopy from IRAS~00420+5530.  The spectra are overplotted with red dotted lines of comparable spectral type from the \citet{Jacoby1984} spectral atlas. \label{spec00420}}
\end{figure}

Figure~\ref{spec00420} shows spectra of the four stars that were obtained revealing two intermediate-mass stars.  Star A has a spectrum with a weak \ion{He}{1} $\lambda$5876 \ang\ absorption feature ($EW$=0.19$\pm$0.02 \AA) and a strong H$\alpha$ emission feature ($EW$=-8.663$\pm$0.02 \AA).  Spectral classification using the \ion{He}{1} line results in a B8e star; however, the  \ion{He}{1} EW may be a lower limit if there there is a nebular emission in that line, therefore, the temperature class could be earlier than B8.  Furthermore, this wavelength region lacks the luminosity diagnostics to distinguish a classical Be star from a B supergiant.  As a result, the spectrophotometric distance for star A was not calculated.  Star B has a spectrum with strong \ion{He}{1} $\lambda$5876 \ang\ absorption ($EW$=0.73$\pm$0.12 \AA) and H$\alpha$ absorption ($EW$=4.491$\pm$0.27 \AA) that appears to indicate an early B star (B2).  Star C has strong, narrow H$\alpha$ absorption ($EW$=6.339$\pm$0.13 \AA) indicative of an early F star (F2).  Star D has a strong, narrow H$\alpha$ absorption ($EW$=4.00$\pm$0.07 \AA) indicative of a late-F star (F9).  We used the spectrum of Star B and its 2MASS colors to derive a spectrophotometric distance of 1.14$^{+0.34+0.47}_{-0.25}$ kpc at an extinction of \av\ = 6.33$^{+0.17}_{-0.17}$ mag for this region.  This distance is not consistent with the maser parallax distance measurement of 2.17 kpc, though the agreement is better than with the kinematic distance measurements of $>$5 kpc.    A portion of this discrepency can be reconciled if Star B is in a binary system with a companion of nearly equal luminosity.

IRAS~00420+5530 contains nine candidate YSOs in the vincinity of the \imsfr.  These include eight intermediate-mass YSOs and one lower-mass YSO.  Of the eight intermediate-mass YSOs, two are Stage I (Y7 and Y12), three are Stage II (Y5, Y10, and Y13), one is Stage III (Y11), and two fits are ambiguous (Y6 lies between Stage I and II and Y9 lies between Stage II and III).  The lower-mass YSO, Y8, is Stage I.   Y6 apparently coincides spatially with Star C.  The spectrum obtained of Star C is consistent with an early F temperature class and shows no indication of accretion.  The SED model fits, however, are consistent with a 5.1 \msun\ YSO.  This may indicate that a foreground F star (estimated distance 0.46 kpc) is projected in front of the IR emission from a intermediate-massive YSO with the \imsfr\ estimated to lie at 2.2 kpc.  In that case, the flux contribution from the YSO and the foreground star would be ambiguous in the 2MASS bands increasing the uncertainty in the YSO fit.  The calculated spectrophotometric distance and extinction listed in Table~\ref{2mass.tbl} is almost certainly in error because of this possible blend.  The total luminosity contribution of these YSOs is 9.6$\times10^2$ \lsun.  Using the \citet{Sanders1996} definition for infrared luminosity, we use the \emph{IRAS} fluxes to derive an infrared luminosity of 8.8$\times10^2$ \lsun\ at 1.14 kpc or 3.2$\times10^3$ \lsun\ at 2.17 kpc.  The total YSO luminosity is consistent with the closer spectro-photometric distance, but if the cluster is at the maser parallax distance would suggest that there are more intermediate-mass YSOs contributing to the infrared luminosity that have not been accounted for.

The cleaned color-magnitude diagram of IRAS~00420+5530 is rather rich, showing a distinct grouping of stars near the nominal main sequence at a distance of 1.14 kpc and reddening of \av\ = 6.33 mag.  Star A lies to the left of the main sequence and is possibly a foregound classical Be star.  Star B, found to be an early B star from our optical spectroscopy, is labeled and lies near the nominal B0V star (upper X) along the main sequence.  Stars C and D (early to late-F) lie to the left of the main sequence. These appear too bright to be mid-F stars at the adopted cluster distance and reddening; they are likely foreground objects at $d\simeq0.26$ kpc and $d\simeq0.71$ (see Table~\ref{2mass.tbl}).  A group of about 25 stars lies near or below the nominal main sequence, most of them between $J-K_S$ = 1.2--1.7. Another sequence of about eight stars stretch redward from pre-main-sequence tracks, consistent with them being either enshrouded pre-main-sequence stars or more reddened background objects.  Two of the nine YSOs identified in the IRAS~00420+5530 field lie within the marked cluster aperture (Y6 and Y12).  Star C (Stage II/III;$\equiv$Y6?) appears on Figure~\ref{cmd1} near \ks = 9.98, $J-K_S\simeq0.6$ and is likely a foreground object along the line of sight toward the YSO, as discussed above.  Y12 (Stage I/II; 5.3 \msun) appears on Figure~\ref{cmd1} near \ks = 9.95, $J-K_S\simeq3.1$ where a young, enshrounded pre-main-sequence star would be expected.  The color-color diagram for IRAS~00420+5530 (Figure~\ref{ccd1}) reveals that the rich grouping of stars from the color-magnitude diagram has a large spread in color from $H-K_S=0.1$ to $H-K_S=0.8$.  There are a few stars stretching along the reddening vector to more extreme colors, including Y12.  Several of these stars appear to have $H-K_S$ colors redder than would be expected for main-sequence objects at higher reddening.  This suggests that these are possible pre-main-sequence stars with k-band excesses.  The nature of the stars that lie to the left of the main sequence is less clear.  The unusually blue $H-K_S$ colors may be the result of reflection nebulosity or poor 2MASS photometry as these stars have typical uncertainties of $\sim$0.2 mag in $J$-, $H$-, and \ks- bands.  Taken together, the cleaned CMD, CCD and the spectra of IRAS~00420+5530 are consistent with a cluster of pre-main-sequence stars in the vinicinity of the early B star that appear to be the most massive constituent of this region.  Using the \citet{Weidner2013} most massive star -- cluster mass relation, we estimate a total cluster mass of $\sim$78 \msun\ in $\sim$250 stars above 0.1 \msun\ after integrating a typical \citet{Kroupa2001} IMF.

\subsection{IRAS~01082+5717 \label{01082.sec}}
IRAS~01082+5717 has largely been unexplored in literature.  It was identified as a site of diffuse \emph{IRAS} emission with a radius of 9\arcmin\ by \citet{Fich1996} in a survey of $IRAS$ 60 \micron\ images.

\begin{figure}
\centering
\includegraphics[scale=0.65,angle=90]{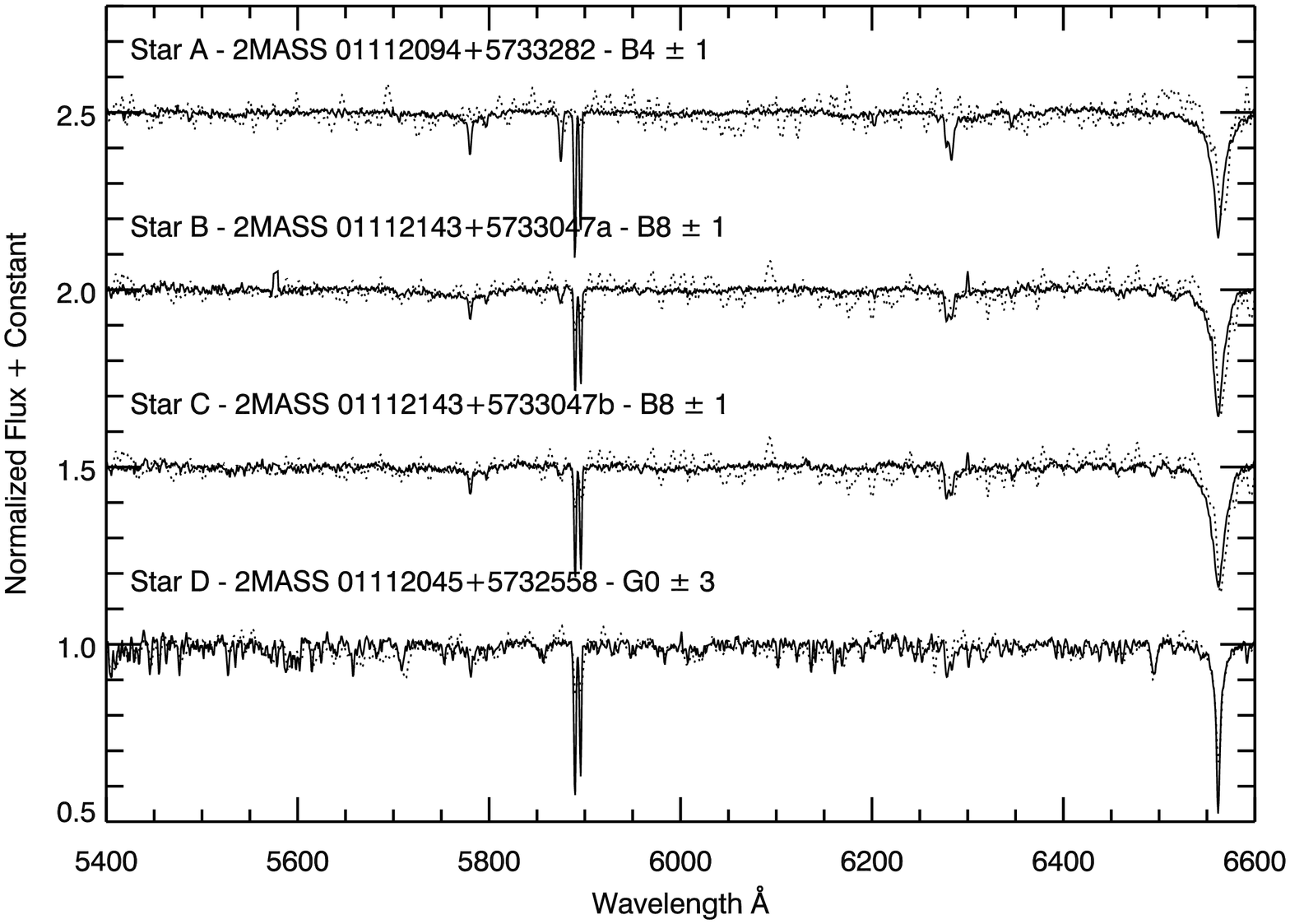}
\caption{Continuum-normalized spectra (solid black lines) of the five stars selected for optical spectroscopy from IRAS01082+5717.  The spectra are overplotted with red dotted lines of comparable spectral type from the \citet{Jacoby1984} spectral atlas. \label{spec01082}}
\end{figure}

Figure~\ref{spec01082} shows the spectra of the four stellar targets revealing three intermediate-mass stars.  Star A has a spectrum with a strong \ion{He}{1} $\lambda$5876 \ang\ absorption feature ($EW$=0.53$\pm$0.01 \AA) and strong H$\alpha$ absorption ($EW$=7.14$\pm$0.52 \AA), indicative of a mid-B star (B4).  Star B shows a weak \ion{He}{1} $\lambda$5876 \ang\ absorption feature ($EW$=0.15$\pm$0.02 \AA) and an H$\alpha$ absorption feature ($EW$=7.40$\pm$0.06 \AA) indicating a late-B star (B8).  Star C exhibits a weak \ion{He}{1} $\lambda$5876 \ang\ absorption feature ($EW$=0.15$\pm$0.03 \AA) and an H$\alpha$ absorption feature ($EW$=8.76$\pm$0.14 \AA) indicating a late-B star (B8).  Star D has a strong, narrow H$\alpha$ absorption feature ($EW$=3.798 $\pm$0.07 \AA) indicative of an early G star (G0).  The separation between Star B and Star C is $\sim$2$\arcsec$ leading to blending in the 2MASS images.  Unfiltered guide camera images taken at the time as the spectroscopic observations indicate that Stars B and C are roughly the same visual magnitude.  The 2MASS catalog photometry for Star B, which includes the flux contribution from Star C, is included in the CMD and CCD for reference, and in Table~\ref{2mass.tbl}.  We used the spectrum of Star A and its 2MASS photometry to derive a spectrophotometric distance of 1.84$^{+0.26+0.76}_{-0.23}$ kpc at an extinction of \av\ = 1.22$^{+0.09}_{-0.11}$ mag for this region.

IRAS~01082+5717 contains five YSOs, all of intermediate mass and all Stage II (Y14, Y15, Y16, Y17, and Y18) in the vincinity of the \imsfr.  Y16--Y18 all lie outside the nominal cluster aperture shown in Figure~\ref{4reg2}, but they are still in the field of IRAS~01082+5717.  Y15 corresponds to Star A.  The spectrum  of Star A presents as a mid-B star ($\sim$5 \msun) at an \av\ of 1.22 mag and shows no indication of accretion.  This agrees with the average SED model fit of a 3.1 \msun\ YSO at an average \av\ of 1.12 mag.  The mean extinctions for the five YSOs vary from 0.86 -- 9.6 mag., but this could be the result of differential extinction within the SFR coupled with the limited precision of the SED fitting technique.  The total luminosity contribution of these YSOs is 2.8$\times10^2$ \lsun.  Using the \citet{Sanders1996} definition for infrared luminosity, we use the \emph{IRAS} fluxes to derive an infrared luminosity of 1.9$\times10^2$ \lsun\ for the region.

The cleaned color-magnitude diagram of IRAS~01082+5717 is rather sparse, showing just a few sources near the nominal main sequence at a distance of 1.84 kpc and reddening of \av\ = 1.22 mag.  Star A, found to be a mid-B star from our optical spectroscopy, is labeled and lies between the nominal B0V and A0V loci (upper and lower X's) along the main sequence.  Star B (late-B) lies above its expected location on the main sequence.  This is likely owing to confusion with Star C (late-B), which was not drawn owing to its lack of photometry in the 2MASS catalog.  Star D (early G) lies just to the right of and above the main sequence consistent with it possibly being a foreground dwarf or  background giant.  Accordingly, we do not compute a spectrophotometric distance for this star.  A group of about seven stars lie along the 2.5 Myr isochrone at locations where pre-main-sequence stars would be expected. Another sequence of about five stars lie along the main-sequence track.  There is also a group of about five stars that stretch redward from pre-main-sequence tracks, consistent with them being either enshrouded pre-main-sequence stars or more reddened background objects.  Two of the five YSOs identified in this field lie within the marked cluster aperture (Y14 and Y15).  Y14 appears on Figure~\ref{cmd2} near $K=13.28$, $J-K_S\simeq1.5$.  Y15 (Star A) lies along the main sequence at the adopted distance and reddening based upon its spectral type.  The color-color diagram  (Figure~\ref{ccd2}) reveals a sparse group of stars near $H-K_S=0.3$, $J-H=0.8$, with a few stretching along the reddening vector to more extreme colors.  Three of these stars appear to have k-band excesses which suggest that they may be pre-main-sequence stars.  Taken together, the cleaned CMD, CCD, and the spectra are consistent with the region hosting a poor cluster of pre-main-sequence stars in the vinicinity of the mid-B and late-B stars that appear to be the most massive constituents.  The limited depth of the 2MASS photometry precludes a more definitive characterization of the possible stellar cluster in IRAS~01082+5717, though using the \citet{Weidner2013} most massive star -- cluster mass relation, we can estimate a total cluster mass of $\sim$37 \msun\ in $\sim$120 stars above 0.1 \msun\ after integrating a typical \citet{Kroupa2001} IMF.

\subsection{IRAS~05380+2020 \label{05380.sec}}
IRAS~05380+2020 is a star-forming region included in an \ion{H}{1} survey by \citet{Lu1990} who found a galaxy with a redshift of 6735 \kms\ at this position.  \citet{Takata1994} argued that the infrared luminosity is very high and not compatible with the \ion{H}{1} profile of a galaxy at such a distance, indicating that there is likely a galaxy behind this Galactic infrared source.

\begin{figure}
\centering
\includegraphics[scale=0.65,angle=90]{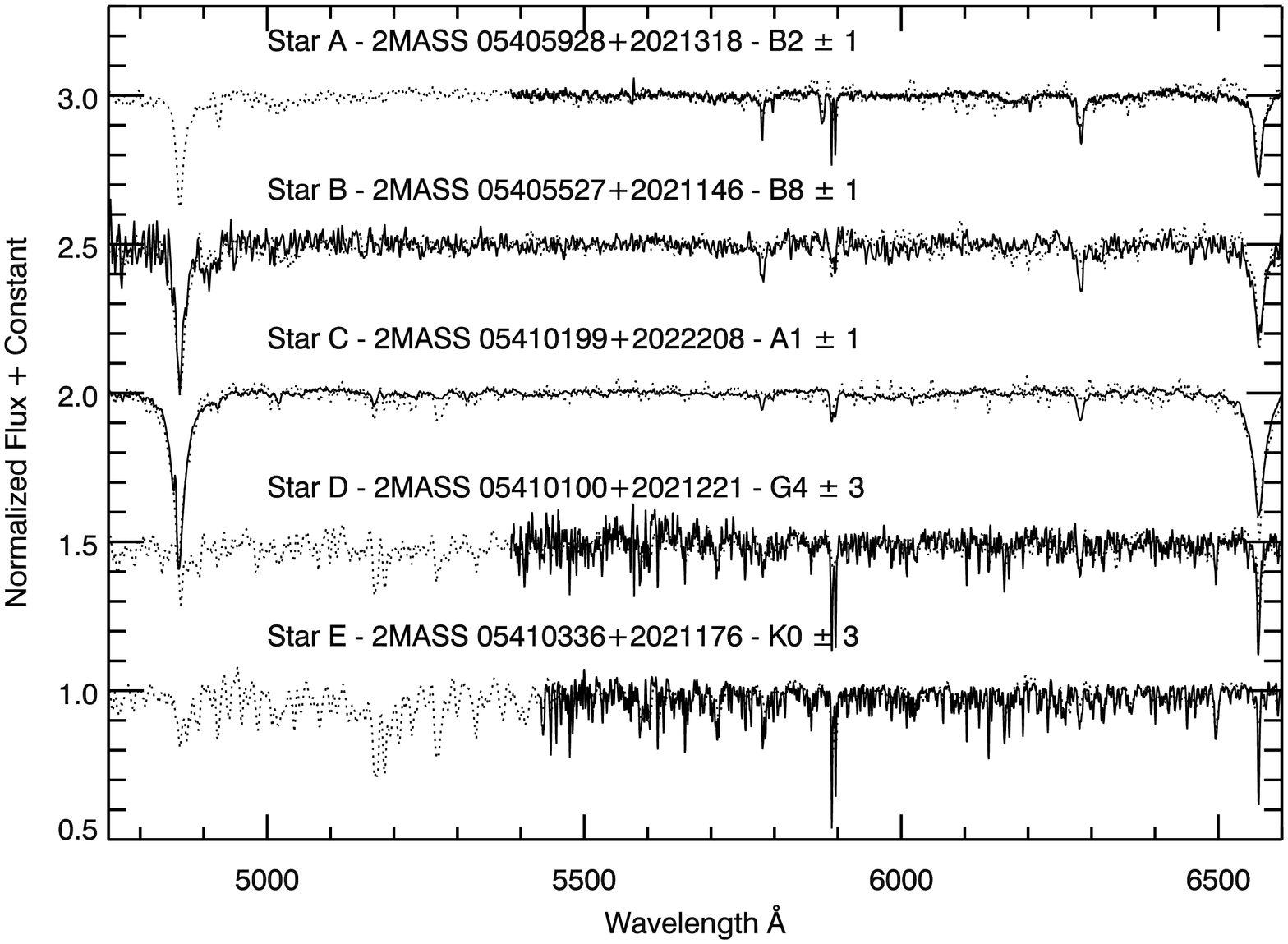}
\caption{Continuum-normalized spectra (solid black lines) of the five stars selected for optical spectroscopy from IRAS05380+2020.  The spectra are overplotted with red dotted lines of comparable spectral type from the \citet{Jacoby1984} spectral atlas. \label{spec05380}}
\end{figure}

Figure~\ref{spec05380} shows the five spectra that were obtained for IRAS~05380+2020, revealing three intermediate-mass stars.  Star A has a strong \ion{He}{1} $\lambda$5876 \ang\ absorption feature ($EW$=0.71$\pm$0.33 \AA) and H$\alpha$ absorption ($EW$=5.61$\pm$0.13 \AA).  This spectrum is indicative of an early B star (B2).  Star B has weak \ion{He}{1} $\lambda$5876 \ang\ absorption ($EW$=0.19$\pm$0.06 \AA) and an H$\alpha$ absorption feature ($EW$=6.63$\pm$0.20 \AA) indicating a late-B (B8) star.  Star C has a strong H$\alpha$ absorption feature ($EW$=14.40$\pm$0.11 \AA) indicating an early A star (A1).  Star D has a narrow H$\alpha$ absorption ($EW$=2.94$\pm$0.24 \AA) indicative of a mid-G (G4) star.  Star E has a  very narrow H$\alpha$ absorption feature ($EW$=1.06$\pm$0.07 \AA) and a strong \ion{Fe}{1} $\lambda$6495 \ang\ absorption feature indicative of an early K (K0) star.  We used the spectrum of Star A and its 2MASS photometry to derive a spectrophotometric distance of 1.34$^{+0.40+0.55}_{-0.30}$ kpc at an extinction of \av\ = 2.74$^{+0.17}_{-0.17}$ mag for this region.

IRAS~05380+2020 contains two candidate YSOs in the vincinity of the \imsfr.  Of these, Y19 is of intermediate mass and Y20 is of lower mass.  Y19 is Stage I and Y20 is ambiguously fit by the models (I/II).  Y19 is poorly fit as a result of a $\sim$5 magnitude increase from the \ks\ band to the \emph{WISE} W1 band.  This may be owing to variability of the source in the $\sim$12 years between the 2MASS and \emph{WISE} observations.  The total luminosity contribution of these YSOs is 8.7$\times10^1$ \lsun.  Using the \citet{Sanders1996} definition for infrared luminosity, we use the \emph{IRAS} fluxes to derive an infrared luminosity of 2.5$\times10^2$ \lsun.

The cleaned color-magnitude diagram of IRAS~05380+2020 is rather rich, showing a distinct grouping of stars near the nominal main sequence at a distance of 1.34 kpc and reddening of \av\ = 2.74 mag.  Star A, found to be an early B star from our optical spectroscopy, is labeled and lies near the nominal B0V locus (upper X) along the main sequence.  Star B, found to be a late-B star, is labeled and lies between the nominal B0V and A0V loci along the main sequence.  Star C lies to the left of the main sequence and above the nominal A0V locus. This appears too bright to be an early A star at the adopted cluster distance and reddening; this is likely a foreground object at $d\simeq0.78$ kpc and correspondingly lower reddening, \av=0.61 mag (see Table~\ref{2mass.tbl}).  A group of about 28 stars scatters about the nominal 2.5 Myr isochrone, most of them between $J-K_S$=1.2 -- 1.6. Another sequence of about six stars stretches redward from the pre-main-sequence tracks, consistent with them being either enshrouded pre-main-sequence stars or more reddened background objects.  One of the two YSOs (Y19)  lies within the marked cluster aperture.  Y19 is located on Figure~\ref{cmd2} near $K$=15.18, $J-K_S$=1.75 where a young, enshrounded pre-main-sequence star would be expected.  The color-color diagram for IRAS~05380+2020 (Figure~\ref{ccd2}) reveals that the rich grouping of stars from the color-magnitude diagram has a small spread in $J-H$ color, but a large spread in $H-K_S$ color from 0.2 to 0.7.  There is also a grouping along the main sequence and several stars stretching along the reddening vector to more extreme colors.  Several stars appear to have k-band excesses suggesting possible pre-main-sequence stars.  Taken together, cleaned CMD, CCD, and the spectra are consistent with a cluster of pre-main-sequence stars in the vinicinity of the early and mid-B stars that appear to be the most massive constituents of this region.  Using the \citet{Weidner2013} most massive star -- cluster mass relation, we estimate a total cluster mass of $\sim$78 \msun\ in $\sim$250 stars above 0.1 \msun\ after integrating a typical \citet{Kroupa2001} IMF.

\subsection{Initial Mass Function Modeling}
\begin{figure}
\centering
\includegraphics[scale=0.6]{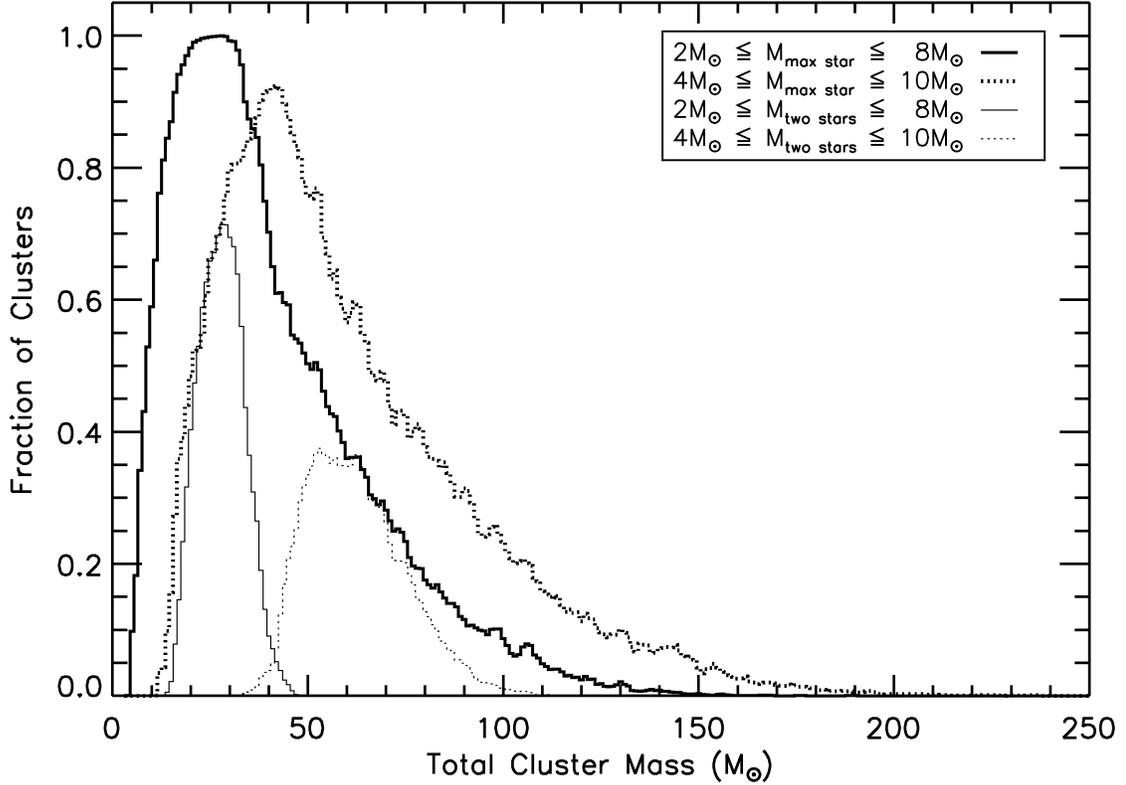}
\caption{Fraction of simulated clusters in each total cluster mass bin that meet the specified stellar population criteria: the most massive star 2\msun\ $\leq$ M$_{\mathrm{max}}$ $\leq$ 8\msun\ (solid bold line), the most massive star 4\msun\ $\leq$ M$_{\mathrm{max}}$ $\leq$ 10\msun\ (dotted bold line), exactly two stars in the mass range 2\msun\ $\leq$ M $\leq$ 8\msun\ including the most massive star (solid light line), and exactly two stars in the mass range 4\msun\ $\leq$ M $\leq$ 10\msun\ including the most massive star (dotted light line).  \label{imf.fig}}
\end{figure}

Given that low- and intermediate- mass clusters suffer from stochastic effects when populating the upper end of the initial mass function, we utilize the stellar cluster simulation software MASSCLEAN \citep{Popescu2009} to model the clusters in our pilot study.  MASSCLEAN allows for stochastic fluctuations in the simulations.  We compute the mass distribution using a Kroupa-Salpeter initial mass function \citep{Kroupa2002,Salpeter1955} with an upper mass limit of 150 \msun.  We randomly generate 1000 clusters for a range of total cluster masses from 3--1000 \msun\ in 1 \msun\ bins and tabulate the fraction that reproduce the upper end of the mass function observed in our clusters.  Figure~\ref{imf.fig} shows the fraction of simulated clusters within each mass bin producing a cluster with the most massive star 2 \msun\ $\leq$ M$_{\mathrm{max}}$ $\leq$ 8 \msun\ (solid bold line) and the most massive star 4 \msun\ $\leq$ M$_{\mathrm{max}}$ $\leq$ 10 \msun\ (dotted bold line).  We also explored requiring the cluster to have exactly two stars in the mass range 2 \msun\ $\leq$ M $\leq$ 8 \msun\ including the most massive star (solid light line) and requiring the cluster to have exactly two stars in the mass range 4 \msun\ $\leq$ M $\leq$ 10 \msun\ including the most massive star (dotted, light line).  The lines in Figure~\ref{imf.fig} have been smoothed to 3 solar mass bins to reduce scatter.  This simulation demonstrates that it is likely that clusters forming only two intermediate mass stars have a total mass between 20 and 80 \msun, consistent with masses estimated from the \citet{Weidner2013} most massive star -- cluster mass relation.  While we cannot completely rule out that these are fairly massive clusters that have happened not to form a massive star, it is highly improbable that clusters with intermediate-mass stars as their most massive constituents have a total mass greater than 150 \msun.  While it is common for a massive cluster to produce a massive star and eject it via few-body interactions to become a runaway OB star, it is highly improbable to dynamically eject the most massive star from a cluster with $<$ 316 \msun \citep{Oh2012}.

\section{Summary and Conclusions\label{conc.sec}}

In this paper we morphologically classify an all-sky sample of 984 candidate \imsfrs\ using \emph{WISE} mid-IR images.  The objects in this all-sky sample were selected from the \emph{IRAS} Point Source Catalog using colors that indicate large PAH contribution and cool dust found in \imsfrs.  We determined that 616 (62.6\%) are classified as ``blobs/shells,'' 128 (13.0\%) are classified as ``filamentary,'' 39 (4.0\%) are classified as ``star-like'' objects, and 201 (20.4\%) are classified as ``galaxies.''  We find that the blobs/shells, filamentary, and star-like objects are concentrated in the Galactic Plane, while galaxies appear to be uniformly distributed throughout the sky.  

We conducted a pilot study  of four Galactic \imsfrs\ that are classified as blobs/shells using optical spectroscopy and near-IR photometry.  Optical spectroscopy of four--five stars per region demonstrates that each contains, within its projected boundaries, two or more intermediate-mass stars.  Eight of the eighteen spectroscopically classified stars are B-type and two are A-type with no stars more massive than $\sim$B2 (8 \msun) found.  Near-IR color-magnitude diagrams using 2MASS data indicate that these stars are the most massive  \& luminous constituents, suggesting that there may be an upper limit to the mass of stars in \imsfrs\ of $\sim$B2.  SEDs of candidate YSOs in each region also indicate that they are continuing to produce intermediate- and lower-mass stars, but not massive stars.  These findings support the idea that the \emph{IRAS} color-selection of (\citealt{Kerton2002,Arvidsson2010}) does indeed select regions producing intermediate-mass stars and that these regions can be distinguished from high-mass SFRs using \iras\ mid-IR colors.

Cleaned 2MASS color-magnitude diagrams reveal a strong excess of stars along the pre-main-sequence isochrones in IRAS~00420+5530 and IRAS~05380+2020 and a less obvious excess in IRAS~00259+5625 and IRAS~01082+5717, indicating the presence of small clusters associated with the spectroscopically identified IM stars.  The former two regions (IRAS~00420+5530 and IRAS~05380+2020) that have the strongest evidence for stellar clusters also the earliest stars as their most massive members ($\sim$B2), while the latter two clusters (IRAS~00259+5625 and IRAS01082+5717) host somewhat later spectral types (B7 and B4, respectively).   Thus, the most massive stars appear to be associated with richer and more massive stellar clusters.   While the current small sample size precludes a definitive statement, this correlation is consistent with the idea that there may be a turn-on point for cluster formation at the cloud masses and stellar masses associated with the formation of mid-B spectral types as the most massive constituent.

\acknowledgements  Acknowledgement:  We would like to acknowledge the support of Carlos A. Vargas Alvarez, Anirban Bhattacharjee, Michael DiPompeo, and Jiajun Chen for their assistance in obtaining spectroscopy data. This work was supported by NASA through grants NAS2-97001 from the Stratospheric Observatory for Infrared Astronomy and ADAP NNX10AD55G.  We thank James Weger and Jerry Bucher for their work in maintaining 
the Wyoming Infrared Observatory as an excellent astronomical facilty. We would also like to thank the anonymous referee for his/her comments and suggestions, helping to improve this manuscript.

\clearpage
\bibliographystyle{apj}
\bibliography{referencelist}

\clearpage

\clearpage

\clearpage

\clearpage

\clearpage

\clearpage

\clearpage

\clearpage

\clearpage

\clearpage

\clearpage

\clearpage

\clearpage

\clearpage

\clearpage

\begin{deluxetable}{ccccc}
\centering
\tabletypesize{\tiny}
\tablewidth{0pc}
\tablecaption{IM~SFR Source Table\label{imsfrsources.tbl}}
\tablehead{
\colhead{Number} &
\colhead{\emph{IRAS} Name} &
\colhead{$\ell$ (\deg)} &
\colhead{$b$ (\deg)} &
\colhead{Morphological Classification} }
\tablecolumns{5}
\startdata

   1 & 00001+6417 &   117.7266 &     2.1899 &    Blob/Shell \\ 
   2 & 00002+6647 &   118.2070 &     4.6387 &   Filamentary \\ 
   3 & 00025+6708 &   118.4940 &     4.9407 &    Blob/Shell \\ 
   4 & 00034+6658 &   118.5597 &     4.7714 &   Filamentary \\ 
   5 & 00063+6706 &   118.8581 &     4.8497 &   Filamentary \\ 
   6 & 00070+6503 &   118.5951 &     2.8161 &    Blob/Shell \\ 
   7 & 00080+6329 &   118.4449 &     1.2530 &    Blob/Shell \\ 
   8 & 00127+6058 &   118.6205 &    -1.3256 &    Blob/Shell \\
\enddata

\end{deluxetable}

\clearpage

\begin{deluxetable}{cccccccc}
\centering
\tabletypesize{\scriptsize}
\tablewidth{0pc}
\tablecaption{Pilot Study \imsfrs \label{sources.tbl}}
\tablehead{
\colhead{\emph{IRAS} Name} &
\colhead{$\ell$} &
\colhead{$b$} &
\colhead{$F_{12}$} &
\colhead{$F_{25}$} &
\colhead{$F_{60}$} &
\colhead{$F_{100}$} &
\colhead{Other ID} \\
\colhead{} &
\colhead{(\deg)} &
\colhead{(\deg)} &
\colhead{(Jy)} &
\colhead{(Jy)} &
\colhead{(Jy)} &
\colhead{(Jy)} &
\colhead{} }
\tablecolumns{8}
\startdata
00259+5625 & 119.8038 & -6.0319 & 0.7603 & 1.116 & 30.56 & 108.5 & CB 3\tablenotemark{a}, LBN 594\tablenotemark{b} \\
00420+5530 & 122.0145 & -7.0718 & 7.328 & 15.13 & 217.0 & 469.7 & Mol 3\tablenotemark{c} \\
01082+5717 & 125.6137 & -5.2090 & 1.138 & 1.089 & 13.56 & 44.78 & --- \\
05380+2020 & 186.7611 & -5.3793 & 2.294 & 1.931 & 34.72 & 118.6 & --- \\
\enddata
\tablenotetext{a}{\citep{Clemens1988}}
\tablenotetext{b}{\citep{LBN}}
\tablenotetext{c}{\citep{Molinari1996}}
\end{deluxetable}

\clearpage
\begin{deluxetable}{ccccccccccccccc}
\centering
\tabletypesize{\tiny}
\setlength{\tabcolsep}{0.04in}
\tablewidth{0pc}
\tablecaption{Spectrophotometric Parameters\label{2mass.tbl}}
\tablehead{
\colhead{\emph{IRAS} Name} &
\colhead{2MASS ID} &
\colhead{J} &
\colhead{H} &
\colhead{\ks} &
\multicolumn{2}{c}{H$\alpha$ 6563} & &
\multicolumn{2}{c}{He \Rom{1} 5876} &
\colhead{Sp. Type} &
\colhead{\av} &
\colhead{D$_{spec}$} &
\colhead{D$_{lit}$} \\

\cline{6-7} \cline{9-10}
\colhead{} &
\colhead{} &
\colhead{} &
\colhead{} &
\colhead{} &
\colhead{EW} &
\colhead{$\sigma$} &
\colhead{} &
\colhead{EW} &
\colhead{$\sigma$} &
\colhead{} &
\colhead{(mag)} &
\colhead{(kpc)} &
\colhead{(kpc)} \\

\colhead{(1)} &
\colhead{(2)} &
\colhead{(3)} &
\colhead{(4)} &
\colhead{(5)} &
\colhead{(6)} &
\colhead{(7)} &
\colhead{(8)} &
\colhead{(9)} &
\colhead{(10)} &
\colhead{(11)} &
\colhead{(12)} &
\colhead{(13)} &
\colhead{(14)} }
\tablecolumns{15}
\startdata
00259+5625 &  &&&&&&&&&&  &  & 2.5\tablenotemark{a}, 2.1\tablenotemark{b}  \\
A & J00283934+5641507 &13.089\tablenotemark{c}&12.896\tablenotemark{c}&12.737& 8.00&0.57 &&0.30 &0.13 & B7 & 2.43$^{+0.07}_{-0.09}$ & 2.12$^{+0.21+0.88}_{-0.21}$   &  \\
B & J00284837+5641468 &13.140&12.960&12.873&14.95 &0.18 &&\nodata &\nodata & A1 & 1.03$^{+0.04}_{-0.03}$&1.65$^{+0.04+0.68}_{-0.08}$ &    \\
C & J00284282+5642554 &12.958&12.632&12.590&4.53 &0.20 &&\nodata &\nodata & F6 & 0.20$^{+0.03}_{-0.04}$&0.91$^{+0.06+0.38}_{-0.06}$  &    \\
D & J00284536+5641525 &12.432&12.088&12.016&3.36 &0.15 &&\nodata &\nodata & G0 & 0.50$^{+0.05}_{-0.04}$&0.59$^{+0.05+0.25}_{-0.05}$ &   \\
E & J00285071+5641430 &11.226&10.512&10.285&1.44 &0.21 &&\nodata &\nodata & G9 & 2.39$^{+0.05}_{-0.41}$&0.19$^{+0.01+0.08}_{-0.05}$ &   \\
00420+5530 &  &&&&&&&&&&  &  & 5.0\tablenotemark{d}, 7.72\tablenotemark{e}, 2.17\tablenotemark{f} \\
A & J00444967+5546225 &10.758&10.292&9.981&-8.66 &0.02 &&0.19 &0.02 & B8e & \nodata&\nodata &  \\
\enddata
\tablecomments{Spectrophotometric distances and \av\ include upper and lower limits resulting from the random uncertainty in spectral type and a second limit {\it upper-only} accounting for systematic uncertainty resulting from the unknown binarity status of each star}
\tablenotetext{a}{Kinematic Distance (\citealt{Launhardt1997})}
\tablenotetext{b}{Kinematic Distance (\citealt{Yun1994a})}
\tablenotetext{c}{Upper Limit}
\tablenotetext{d}{Kinematic Distance (\citealt{Molinari2002})}
\tablenotetext{e}{Kinematic Distance (\citealt{Zhang2005})}
\tablenotetext{f}{Maser Trigonometric Parallax (\citealt{Moellenbrock2009})}
\tablenotetext{g}{Foreground F star and background YSO both contributing flux}
\tablenotetext{h}{Blended in 2MASS}


\end{deluxetable}
\clearpage

\begin{deluxetable}{cccccccccccc}
\centering
\tabletypesize{\tiny}
\tablewidth{0pc}
\setlength{\tabcolsep}{0.04in}
\tablecaption{Candidate YSOs}
\tablehead{
\colhead{\emph{IRAS} Name} &
\colhead{YSO ID} &
\colhead{WISE ID} &
\colhead{J} &
\colhead{H} &
\colhead{\ks} & &
\colhead{W1} &
\colhead{W2} & 
\colhead{W3} &
\colhead{W4} }

\tablecolumns{11}
\startdata
00259+5625&&&&&&&&&&&\\
& Y1 & J002841.65+564152.7   &   17.111 & 14.439 & 12.371 & & 10.460 & 9.262 &  6.667 & 3.736 &   \\
& Y2 & J002831.33+564419.9   &  11.998 & 11.770 & 11.656 & & 11.410 & 11.281 &  7.827 & 5.799 &    \\
& Y3 & J002845.22+564307.6  &   17.233 & 14.718 & 12.744 & & 11.271 & 10.181 & 8.412   & 6.298 &    \\
& Y4 & J002836.86+564245.4  &   16.267 & 14.490 & 13.672 & & 12.564 & 11.858 &  7.134 & 4.341 &    \\
00420+5530&&&&&&&&&&&\\
& Y5 & J004356.01+555118.6  &   15.169 & 14.265 & 13.951 & & 12.255 & 11.651 &   9.595 & 7.661 &    \\
& Y6 & J004503.67+554648.9   &  10.564 & 10.362 & 9.979 & & 8.693 & 7.980 &  4.345 & 1.532 &    \\
\enddata
\label{ysoinput.tbl}
\end{deluxetable}
\clearpage

\begin{deluxetable}{ccccccccccccccccccccccc}
\centering
\tabletypesize{\tiny}
\tablewidth{0pc}
\setlength{\tabcolsep}{0.02in}
\tablecaption{Candidate YSO Fitted Parameters}
\tablehead{
\colhead{\emph{IRAS} Name} &
\colhead{ID} &
\colhead{\# of Fits} &
\colhead{$\chi^{2}/\mathrm{n}_{\mathrm{data}}$} &
\multicolumn{2}{c}{\av} & &
\multicolumn{2}{c}{$\mathrm{M}_{*}(\mathrm{M}_\odot)$} & &
\colhead{$\mathrm{M}_{*}(\mathrm{M}_\odot)$} &
\multicolumn{2}{c}{$\mathrm{log}_{10}(\mathrm{M}_{\mathrm{disk}}/\mathrm{M}_\odot)$}  & &
\multicolumn{2}{c}{$\mathrm{log}_{10}(\dot{\mathrm{M}}/\mathrm{M}_\odot \mathrm{yr}^{-1})$} & &
\colhead{$\mathrm{log}_{10}(\mathrm{L}/\mathrm{L}_\odot)$} &
\multicolumn{2}{c}{Integer Stage} &
\colhead{$\mathrm{F}_{\mathrm{stage}}$} &
\colhead{$\sigma_{\mathrm{stage}}$} \\
\cline{5-6} \cline{8-9} \cline{12-13} \cline{15-16} \cline{19-20} \\

\colhead{} &
\colhead{} &
\colhead{} &
\colhead{BEST} &
\colhead{BEST} &
\colhead{AVG} & &
\colhead{BEST} &
\colhead{AVG} & &
\colhead{Range} &
\colhead{BEST} &
\colhead{AVG} & &
\colhead{BEST} &
\colhead{AVG} & &
\colhead{AVG} &
\colhead{BEST} &
\colhead{AVG} &
\colhead{AVG} &
\colhead{AVG} }
\tablecolumns{21}
\startdata
00259+5625&&&&&&&&&&&&&&&&&&&&&&\\
& Y1 & 294 &   0.03 & 24.72 & 24.09 & &3.7 & 4.0 & &1.5--5.7&-3.22 & -1.41 & & -4.94 & -4.37 & &2.14&   I & amb & 1.54 & 0.53 \\
& Y2 & 163 &   0.94 &  0.00 &  0.42 & &2.3 & 2.8 & &0.8--3.7&-1.70 & -2.55 & & -7.12 & -6.32 & &1.39&  II &  II & 2.07 & 0.19 \\
& Y3 & 530 &   0.87 & 16.38 & 17.56 & &3.0 & 3.0 & &1.0--4.5&-1.31 & -1.99 & &  0.00 & -6.26 & &1.82&  II &  II & 1.99 & 0.02 \\
& Y4 &  51 &   0.50 &  4.83 &  4.86 & &1.2 & 0.8 & &0.3--3.0&-1.62 & -1.84 & & -4.91 & -4.52 & &1.00&   I &   I & 1.02 & 0.04 \\
00420+5530&&&&&&&&&&&&&&&&&&&&&&\\
& Y5 &  41 &   2.73 &  5.63 &  5.18 & &1.9 & 1.9 & &0.1--2.8&-5.07 & -2.95 & &  0.00 & -7.56 & &1.12&  II &  II & 1.99 & 0.02 \\
& Y6 &  24 &   0.87 &  2.27 &  2.44 & &5.2 & 5.1 & &4.8--5.2&-5.83 & -5.54 & &  0.00 &  0.00 & &2.70& III & amb & 2.66 & 0.45 \\
\enddata
\label{newysoresults.tbl}
\end{deluxetable}
\clearpage

\end{document}